\documentclass[pra, twocolumn,showpacs,nofootinbibfloatfix,amsmath,amsfonts,amssymb]{revtex4-1}%
\usepackage{amsmath,amsfonts,amssymb,color}
\usepackage{amsthm}
\usepackage{leftidx}
\usepackage{graphicx}
\usepackage{xcolor}
\usepackage{dcolumn}
\usepackage{bm}
\usepackage{epstopdf}
\usepackage{epsfig}

\newcommand{\n}{\noindent}

\begin{document}
	\title{Measurement-only quantum computation with Majorana corner modes}
	\author{Raditya Weda Bomantara}
	\email{phyrwb@nus.edu.sg}
	\affiliation{%
		Department of Physics, National University of Singapore, Singapore 117543
	}
	\author{Jiangbin Gong}%
	\email{phygj@nus.edu.sg}
	\affiliation{%
		Department of Physics, National University of Singapore, Singapore 117543
	}
	\date{\today}
	
	
	\vspace{2cm}

	\begin{abstract}
		
		Majorana modes, typically arising at the edges of one-dimensional topological superconductors, are considered to be a promising candidate for encoding nonlocal qubits in fault-tolerant quantum computation. Here we propose to exploit the two-dimensional geometry of Majorana corner modes  in second-order topological superconductors to realize measurement-only quantum computation.  In particular,
		a periodically modulated topological superconducting system  may host eight Majorana corner modes, through which two logical qubits as well as one ancilla qubit can be constructed and topologically protected gate operations can be implemented by a series of Majorana parity measurements. The measurements are achieved via Mach$-$Zehnder type interference in the conductance of a semiconductor based topological superconductor.  Our proposal represents a scenario in which single- and {two-qubit} gate operations can be carried out in a minimal setup that may provide a natural building block for Majorana-based qubit architectures.
		
	\end{abstract}
	
	\maketitle
	
	\textit{Introduction.} In recent years, topological quantum computing has emerged as a new paradigm towards establishing fault-tolerance and stimulated many theoretical and experimental studies \cite{tqc,tqc2}. Promising candidates for this purpose are the so-called Majorana fermions, which may arise as zero energy excitations at the vortices or edges of topological superconductors \cite{Ivanov,Kit} and can form nonlocal qubits. A number of experimental signatures evidencing such exotic quasiparticles have been established over the years \cite{exp1,exp2,exp3}, and current efforts have shifted towards demonstrating their ability to store and process quantum information.
	
	Manipulating Majorana qubits is accomplished through a process called braiding, which amounts to moving a pair of Majorana fermions around each other. While various braiding protocols have been proposed over the years \cite{braid1,braid2,braid3,braid4,braid5,braid6,braid7,braid8}, several challenges that hinder their experimental realizations have yet to be solved. These include the necessity to introduce complicated geometries to facilitate braiding, the large space overhead for hosting a number of Majorana fermions, and the large time overhead for completing a single braiding process via adiabatic evolution.
	
	In this paper, we propose the use of the recently discovered higher-order topological phases \cite{HTI1,HTI2,HTI3,HTI4,HTI5,HTI6,HTI7,HTI8,HTI9,HTI10,HTI11,HTI12,HTI13,HTI14,HTI15,HTI16,HTI17,HTI18,HTI19,HTI20,HTI21,HTI22,HTI23,HTI24} for Majorana-based quantum computing. Indeed, a number of recent studies on second-order topological superconductors (SOTSCs) have touched upon Majorana modes localized at the corners of a two-dimensional (2D) system \cite{HTI7,HTI8,HTI9,HTI10,HTI11,HTI12}, termed Majorana corner modes. The main task of this work is to uncover the potential of corner Majorana modes in quantum computing.   As shown below, the 2D geometry of such SOTSC {makes it possible for us to further develop explicit measurement-based quantum gate operations~\cite{mea,mea1,mea2,mea3,mea4,mea5,mea6}.  Indeed, this theoretical work indicates new possibilities in significantly minimizing the time overhead and in overcoming a few other challenges} usually associated with physical braiding processes, such as the requirement for complicated designs and the accumulation of nonlocal errors \cite{err1,err2}.
	
	With periodic modulation in the system parameters, the resulting Floquet SOTSCs may host two different species of Majorana corner modes at quasienergy zero and $\pi/T$ ($T$ being the driving period), which correspond to Majorana zero modes (MZMs) and Majorana $\pi$ modes (MPMs) respectively \cite{FMF1,FMF2,FMF3,braid6,braid7}. Utilizing all these Majorana modes allows the encoding of three qubits. The system's 2D geometry also naturally facilitates Mach$-$Zehnder interferometer like conductance measurements to read out the joint parities of two and four Majorana operators.  These suffice to implement all single- and {two-qubit} Clifford gate operations, thus unleashing the full potential of Majorana qubits in a minimal setup. In addition, scaling up our approach is expected to require significantly less space overhead as compared with many existing qubit architecture proposals.
	
	\textit{Minimal model.} A Floquet SOTSC with four MZMs and MPMs at its corners may be described by a periodically driven $p_x+\mathrm{i}p_y$ superconductor with dimerized pairing in the $y$-direction, as described by the Hamiltonian,
	
	\begin{eqnarray}
	H_s(t) &=&\sum_{i,j} \left[-J_{y,j} c_{i,j+1}^\dagger c_{i,j}  \right. -J_x c_{i+1,j}^\dagger c_{i,j} + \frac{\mu_j(t)}{2} c_{i,j}^\dagger c_{i,j}\nonumber \\
	&& \left. +\mathrm{i} \Delta_{y,j} c_{i,j+1}^\dagger c_{i,j}^\dagger +\Delta_x c_{i+1,j}^\dagger c_{i,j}^\dagger +h.c. \right]\;,
	\label{mod}
	\end{eqnarray}
	
	\n where $J_{y,j}=J_y+(-1)^j \delta J$ and $\Delta_{y,j}=\Delta_y +(-1)^j \delta_y$ are the hopping and pairing amplitudes in the $y$-direction, $\mu_j(t)=\mu_0+(-1)^j \delta \mu_0+[\mu_1 +(-1)^j \delta \mu_1]\cos(\omega t)$ is the periodically driven chemical potential with period $T=\frac{2\pi}{\omega}$, $c_{i,j}$ ($c_{i,j}^\dagger$) is the annihilation (creation) operator at lattice site $(i,j)$, and $2N_x$ ($2N_y$) is the lattice size in the $x$-($y$-)direction. To partially digest this model construction, we consider the special case of $J_y=\delta J$ and $\Delta_y=\delta_y$, where the sites with $j=1, 2N_y$ are decoupled from the rest and the system at $j=1, 2N_y$  effectively reduces to a periodically driven one-dimensional (1D) Kitaev chain \cite{FMF1,FMF2,FMF3,braid6,braid7}. By defining two species of Majorana operators $\gamma_{i,j,A}=c_{i,j}+c_{i,j}^\dagger$ and $\gamma_{i,j,B}=\mathrm{i}\left(c_{i,j}-c_{i,j}^\dagger\right)$, the effective one-dimensional (1D) model at $j=1, 2N_y$ reads
	
	\begin{equation}
	H_{\rm 1D}(t) = \mathrm{i}\sum_{i}\left[J \gamma_{i+1,j,A}\gamma_{i,j,B}+\mu(t) \gamma_{i,j,B} \gamma_{i,j,A} \right] \;,
	\label{mod1d}
	\end{equation}
	
	\n where $\Delta_x=J_x=J$ and $j=1,2N_y$. It follows that if $H_{\rm 1D}(t)$ is tuned to host one pair of MZMs and one pair of MPMs, the full two-dimensional (2D) system will then host eight Majorana corner modes in total. Due to their topology protection {relying on particle-hole symmetry alone}, these corner modes continue to exist for a range of other parameter values with $J_y\neq \delta J$ and $\Delta_y\neq \delta_y$.  Additional properties of Eq.~(\ref{mod}), such as its symmetry analysis and further computational results of the corner modes, are presented in the Supplemental Material \cite{Supp}.
	
\begin{figure}
		\begin{center}
			\includegraphics[scale=0.4]{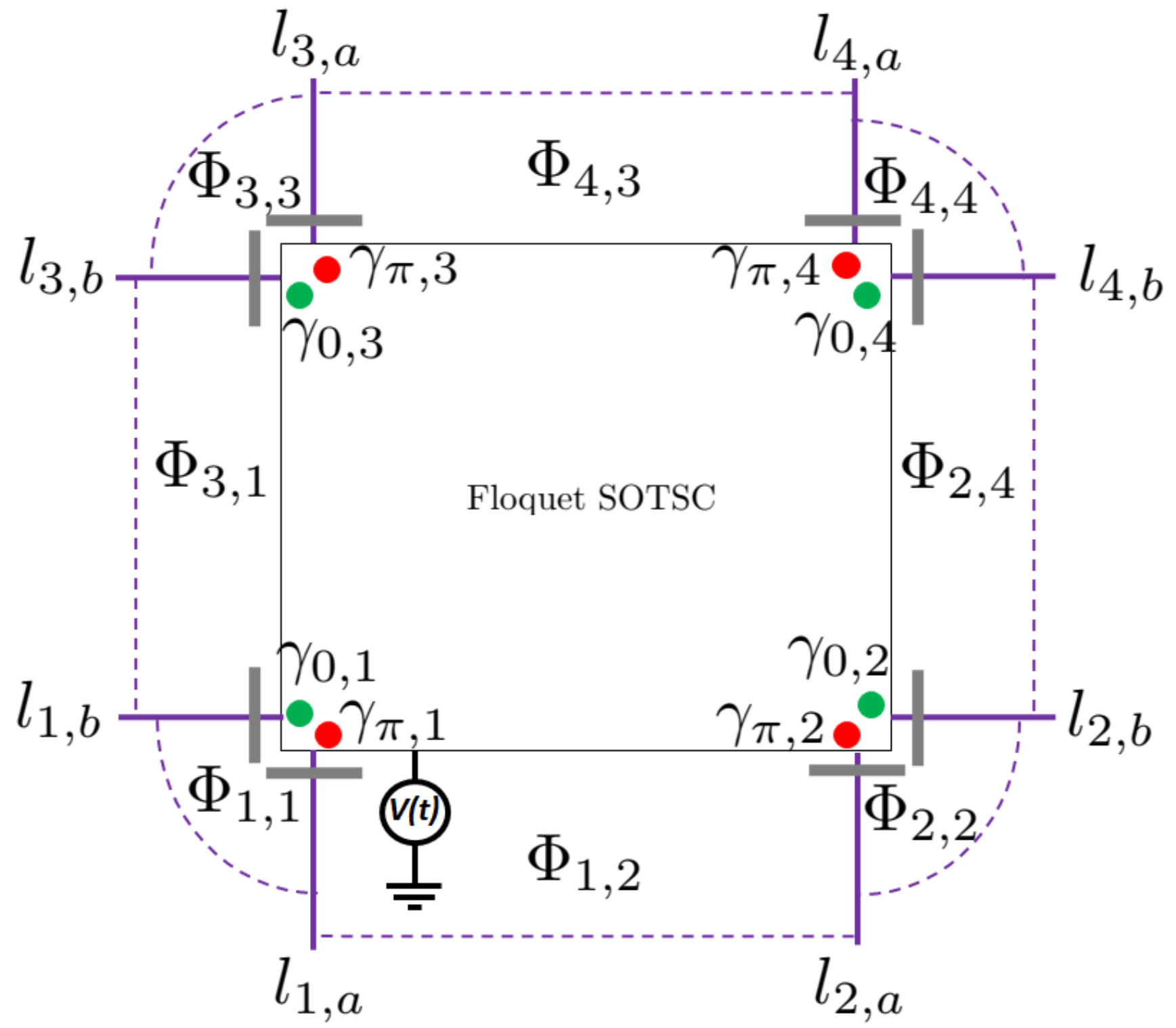}
		\end{center}
		\caption{Schematics of Floquet SOTSC hosting eight Majorana modes connected to two leads at each of its corner. Each lead is connected to its adjacent ones through a weak link marked in dashed purple.}
		\label{lead}
	\end{figure}

	In the following we label the four MZMs and MPMs in a single Floquet SOTSC by $(0,j)$ and $(\pi,j)$ shown in Fig.~\ref{lead}, where  $j=1,2,3,4$. In the second-quantization language, Floquet MZMs and MPMs are Hermitian time-periodic operators $\gamma_{0,j}$ (of period $T$) and $\gamma_{\pi,j}$ (of period $2T$),  which satisfy the commutation relations $\left[H-\mathrm{i}\hbar\frac{d}{dt},\gamma_{0,j}\right]=0$ and $\left[H-\mathrm{i}\hbar\frac{d}{dt},\exp(\mathrm{i}\omega t/2)\gamma_{\pi,j}\right]=\frac{\hbar \omega}{2}\exp(\mathrm{i}\omega t/2)\gamma_{\pi,j}$ \cite{Supp}. In terms of $\gamma_{0,j}$ and $\gamma_{\pi,j}$, three topological qubits can then be defined as $\sigma_z^{(1)}=\mathrm{i} \gamma_{0,1} \gamma_{0,2}$, $\sigma_x^{(1)}=\mathrm{i} \gamma_{0,1} \gamma_{0,3}$, $\sigma_z^{(2)}=\mathrm{i} \gamma_{\pi,1} \gamma_{\pi,2}$, $\sigma_x^{(2)}=\mathrm{i} \gamma_{\pi,1} \gamma_{\pi,3}$, $\sigma_z^{(3)}=\gamma_{0,1} \gamma_{0,2}\gamma_{0,3} \gamma_{0,4}$, and $\sigma_x^{(3)}=\mathrm{i} \gamma_{0,4}\gamma_{\pi,4}$. The parities of these qubit operators depict the presence/absence of nonlocal fermions. For example, eigenvalue $+1$ of $\sigma_z^{(2)}$ indicates the presence of a nonlocal fermion composed from MPMs at corners $j=1,2$; whereas parity $+1$ of $\sigma_z^{(3)}$ indicates the simultaneous presence or absence of two nonlocal fermions composed from MZMs at corners $j=1,2,3,4$ \cite{note}. While $\gamma_{0,j}$, $\gamma_{\pi,j}$, and thus the qubits defined above are all time dependent, their quantum expectation values on the system's time evolving subspace of MZMs and MPMs are constant by construction, thus allowing quantum information to be stored at any moment in time. We shall elaborate below a means to experimentally measure such qubits, which can in turn be utilized for measurement-only quantum computation.

	\textit{Measurement-based computation.}
	Figure~\ref{lead} shows a semiconducting-based \cite{sc1,sc2} Floquet SOTSC, which is coupled to an external capacitor that adds an additional term $H_C(t)=E_C(\hat{N}-N_0(t))$ in the Hamiltonian, where $E_C=\frac{e^2}{2C}$ is the charging energy, $C$ is the capacitance, and the offset $N_0(t)$ is controlled by the gate voltage across the capacitor, which can be static or time-periodic with the same period $T$ as the system. In addition, two leads are attached to each corner of the system and each lead-system coupling can be individually switched on and off on-demand via external gates (marked in gray). For simplicity, each lead is assumed to be a single-level system with energy $E_i=\frac{n_i \hbar \omega}{2}$ {at zero temperature}, where $i=l_{s,m}$ is the lead index shown in Fig.~\ref{lead}, $s=1,2,3,4$, $m=a,b$, and $n_i\in \mathbb{Z}$. Any two adjacent leads are further weakly coupled with each other (independent of the system), as marked by the dotted lines in Fig.~\ref{lead}, which encloses a tunable magnetic flux $\Phi_{s,s'}$. The results presented below are expected to hold also for multi-level leads with other energy levels being sufficiently far away from $\frac{n\hbar \omega}{2}$, where $n$ is an integer.
	
	In the Coulomb blockade regime, i.e., when $E_C$ is sufficiently large (which is still smaller than the quasienergy gaps around zero and $\pi/T$ quasienergies), the transport mechanism between Floquet SOTSC and external leads is dominated by the eigenstates closest to quasienergy zero or $\pi/T$ with particle numbers $N$ and $N\pm 1$.  This accounts for processes in which only a single particle moves in and out the system. Moreover, such eigenstates depend on the lead energies $E_i$ and $E_j$, such that by adjusting the integers $n_i$ and $n_j$, Floquet perturbation theory involving only these states then takes us to an effective Hamiltonian that is dominated by the desired product of Majorana modes.
	
	To further elaborate the above argument, we start by switching on any two lead-system couplings corresponding to leads $i=l_{s,m}$ and $j=l_{s',m'}$. The total lead-system Hamiltonian can then be written as
	
	\begin{eqnarray}
	H_{\rm tot} &\approx & \left\lbrace \sum_{s=i,j}  d_s^\dagger \left[\lambda_{0,s}(t) \gamma_{0,s}(t) + \lambda_{\pi,s}(t) \gamma_{\pi,s}(t)\right] e^{-\mathrm{i}\phi} \right. \nonumber \\
	&& \left.+\lambda_{i,j} e^{\mathrm{i} e\Phi_{i,j}/\hbar} d_j^\dagger d_i+ E_i d_i^\dagger d_i + E_j d_j^\dagger d_j +h.c.\right\rbrace   \nonumber \\
	&& +H_C(t)+H_s(t) \;,
	\label{htot}
	\end{eqnarray}
	where $d_s$ ($d_s^\dagger$) is the lead $s$ annihilation (creation) operator, $\lambda_{i,j},\lambda_{0(\pi),s}\ll E_C$ are respectively the coupling strength between the two leads and between lead $s$ and the MZM $\gamma_{0,s}$ (MPM $\gamma_{\pi,s}$), and $e^{\pm\mathrm{i}\phi}$ is the particle raising/lowering operator which satisfies $[\hat{N},e^{\pm\mathrm{i}\phi}]=\pm e^{\pm\mathrm{i}\phi}$. Note that the actual
	coupling between the leads and the end sites of the wire can be assumed to be static. The time dependence
	in the leads-MZM(MPM) coupling coefficients $\lambda_{0,s}(t)$ and $\lambda_{\pi,s}(t)$ in Eq.~(\ref{htot}) emerges only because the used representations MZMs and MPMs in Eq.~(\ref{htot}) are themselves time periodic operators.  It is thus obvious that the Fourier coefficients $\bar{\lambda}_{0(\pi),s,n}=\frac{1}{T}\int_0^{T} dt \lambda_{0(\pi),s}e^{-\mathrm{i} n \omega t/2}$ will be most pronounced for $n=0$ or
	$n=\pm 1$.
	

	\begin{figure}
		\begin{center}
			\includegraphics[scale=0.45]{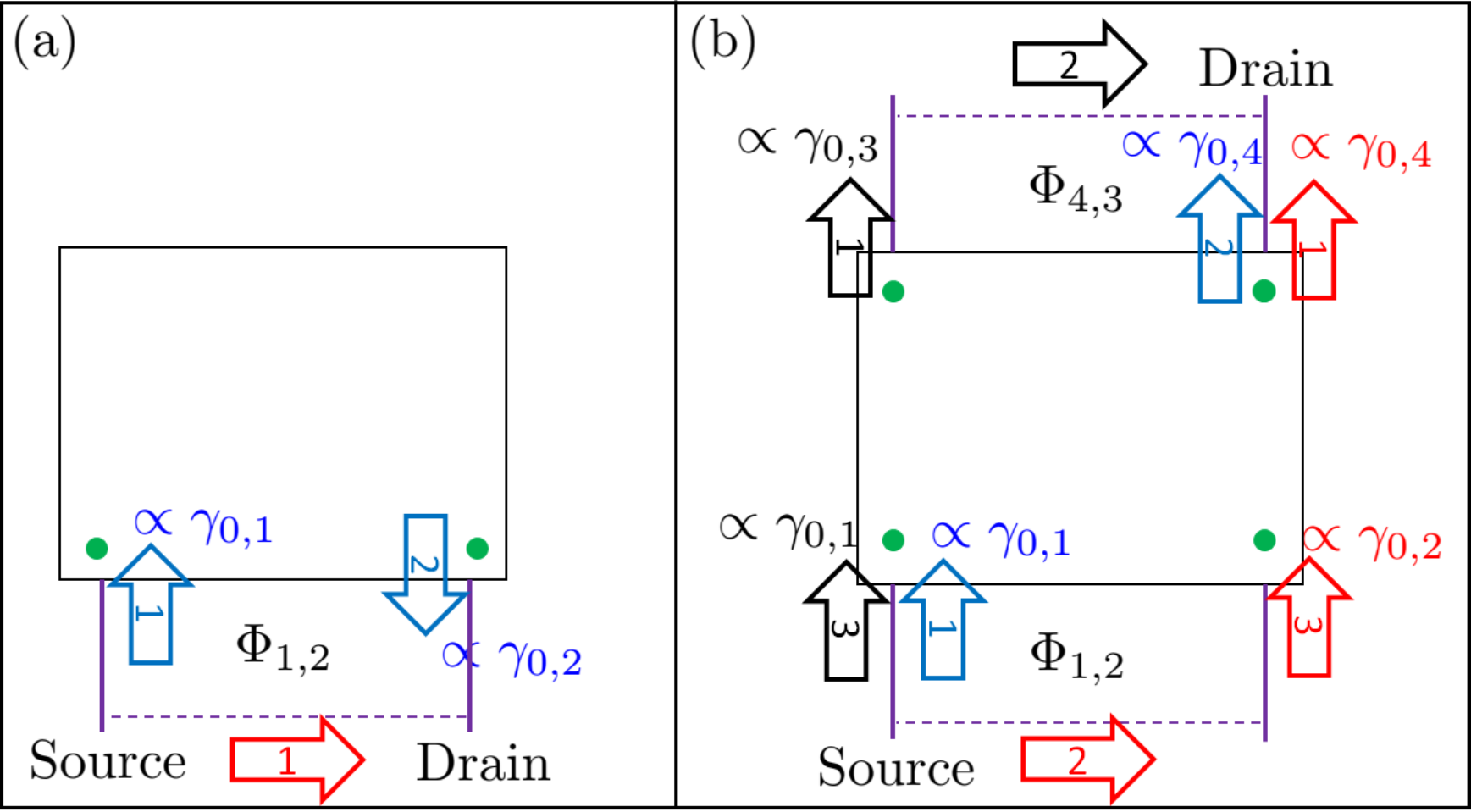}
		\end{center}
		\caption{Interference mechanism that allows the measurements of (a) $\mathrm{i}\gamma_{0,1}\gamma_{0,2}$ and (b) $\gamma_{0,1}\gamma_{0,2}\gamma_{0,3}\gamma_{0,4}$. All leads are tuned to zero energy, conductance measurement is performed between source and drain leads, and all possible particle paths (up to third order process) between source and drain leads are highlighted by different colored arrows. Particle entering or leaving the system will pick up a term that depends on a Majorana corner mode.}
		\label{lead2}
	\end{figure}
	
	By treating the leads-system coupling as a perturbation, it can be effectively replaced (up to second order in $\lambda_{0(\pi),s}$) by a parity-dependent effective Hamiltonian given by one of the following expressions depending on the lead energies $E_i=\frac{n_i\hbar \omega}{2}$ and $E_j=\frac{n_j \hbar \omega}{2}$ \cite{Supp},
	
	\begin{eqnarray}
	H_{\rm eff}^{(00)} &=&  T_{i,j}^{(00)}(t) \mathrm{i} \gamma_{0,i} \gamma_{0,j} d_j^\dagger d_i +h.c. \;, \nonumber \\
	H_{\rm eff}^{(\pi\pi)} &=& T_{i,j}^{(\pi\pi)}(t) \mathrm{i} \gamma_{\pi,i} \gamma_{\pi,j} d_j^\dagger d_i +h.c. \;, \nonumber \\
	H_{\rm eff}^{(0\pi)} &=& T_{i,j}^{(0\pi)}(t) \mathrm{i} \gamma_{0,i} \gamma_{\pi,j} d_j^\dagger d_i +h.c. \;. \label{pert}
	\end{eqnarray}
	As detailed in the Supplemental Material \cite{Supp}, $H_{\rm eff}^{(00)}$ ($H_{\rm eff}^{(\pi\pi)}$) is obtained by setting both $n_i$ and $n_j$ to be even (odd) integers, while $H_{\rm eff}^{(0\pi)}$ is obtained by setting $n_i$ and $n_j$ to be even and odd respectively. The tunneling amplitudes $T_{i,j}^{(00)}(t)$, $T_{i,j}^{(\pi\pi)}(t)$, and $T_{i,j}^{(\pi\pi)}(t)$, generally depend on the Fourier components of the lead-Majorana couplings $\bar{\lambda}_{0(\pi),s,n}$ ($n=0, \pm 1$) and the quasienergy difference between states with $N$ and $N\pm 1$ particles ($\varepsilon_\pm = \varepsilon_N-\varepsilon_{N\pm 1}$).

	Physically, Eq.~(\ref{pert}) can also be understood as a co-tunneling process in which one particle from lead $i=l_{s,m}$ enters the system while another particle leaves the system through lead $j=l_{s',m'}$, mediated by a nonlocal fermion formed by two Majorana corner modes near leads $i$ and $j$ (see Fig.~\ref{lead2}(a)). By assuming that such a co-tunneling process makes a comparable contribution as that of the direct lead coupling (that is, $\left|\lambda_{i,j}\right| \sim \left|T_{i,j}^{(00)}\right|,\left|T_{i,j}^{(\pi\pi)}\right|,\left|T_{i,j}^{(0\pi)}\right|$), their interference can be appreciable. { To further control the strength of the interference term after considering the various time dependence of $T_{i,j}^{(00)}$, $T_{i,j}^{(\pi\pi)}$, and $T_{i,j}^{(0\pi)}$, we may take a time periodic $\Phi_{i,j}(t)=\Phi_{i,j,0}+\Phi_{i,j,1}\sin(\omega t)$}, so that
	the total ($T$-averaged) conductance between the two leads is given by (each associated with one of Eq.~(\ref{pert})) \cite{mea6}
	
	\begin{eqnarray}
	\bar{G}^{(00)}_{i,j} &\propto & \frac{1}{T} \int_0^{T} dt \left|T_{i,j}^{(00)}(t) \langle  \mathrm{i} \gamma_{0,i}\gamma_{0,j}\rangle  +\lambda_{i,j} e^{\mathrm{i}e\Phi_{i,j}/\hbar} \right|^2 \nonumber \\
	&=& g_0^{(00)} +g_1^{(00)} \langle  \mathrm{i} \gamma_{0,i}\gamma_{0,j}\rangle \sin\left[\frac{e}{\hbar}\left(\Phi_{i,j,0}-\Phi_{i,j}^{(00)}\right)\right] \;, \nonumber \\
	\bar{G}^{(\pi\pi)}_{i,j} &\propto & \frac{1}{T} \int_0^{T} dt \left|T_{i,j}^{(\pi\pi)}(t)\langle \mathrm{i}   \gamma_{\pi,i}\gamma_{\pi,j}\rangle +\lambda_{i,j} e^{\mathrm{i} e\Phi_{i,j}/\hbar} \right|^2 \nonumber \\
	&=& g_0^{(\pi\pi)} +g_1^{(\pi\pi)} \langle  \mathrm{i} \gamma_{\pi,i}\gamma_{\pi,j}\rangle \sin\left[\frac{e}{\hbar}\left(\Phi_{i,j,0}-\Phi_{i,j}^{(\pi\pi)}\right)\right] \;, \nonumber \\
	\bar{G}^{(\pi 0)}_{i,j} &\propto & \frac{1}{T} \int_0^{T} dt \left|T_{i,j}^{(0\pi)}(t) \langle \mathrm{i} \gamma_{0,i}\gamma_{\pi,j} \rangle+\lambda_{i,j} e^{\mathrm{i} e\Phi_{i,j}/\hbar} \right|^2 \nonumber \\
	&=& g_0^{(\pi\pi)} +g_1^{(\pi\pi)} \langle  \mathrm{i} \gamma_{\pi,i}\gamma_{\pi,j}\rangle \sin\left[\frac{e}{\hbar}\left(\Phi_{i,j,0}-\Phi_{i,j}^{(\pi\pi)}\right)\right] \;, \nonumber \\
	\label{cond1}
	\end{eqnarray}
	where $g_0^{(ll')}$, $g_1^{(ll')}$, and $\Phi_{i,j}^{(ll')}$, with $l,l'=0,\pi$, are all constants that depend on the system parameters such as $\bar{\lambda}_{l,s,n}$, $\lambda_{i,j}$, and $\varepsilon_\pm$. { In particular, the interference constant $g_1^{(ll')}\propto J_{n/2}(e\Phi_{i,j,1}/\hbar)$ can be controlled via the parameter $\Phi_{i,j,1}$, where $n\in\mathbb{Z}$ depends on the lead energies and $J_\nu(x)$ is the Bessel function of the first kind.} Since $\langle \mathrm{i} \gamma_{l,i} \gamma_{l',j}\rangle$ with $l,l'=0,\pi$ is independent of time and gives either $\pm 1$ depending its the parity, measuring $\bar{G}^{00}_{i,j}$, $\bar{G}^{\pi\pi}_{i,j}$, or $\bar{G}^{0\pi}_{i,j}$ thus fulfills the measurement of any Majorana parity $\mathrm{i} \gamma_{l,i} \gamma_{l',j}$.
	
	In a similar fashion, joint parity measurements of four Majorana modes can be carried out by switching on appropriate four lead-system couplings and measuring the conductance between two of the leads. In particular, by switching on leads $l_{1,a}$, $l_{2,a}$, $l_{3,a}$, and $l_{4,a}$ and setting their energies to zero, the $T$-averaged conductance between leads $l_{1,a}$ and $l_{4,a}$ can be found as $\bar{G}_{l_{1,a},l_{4,a}}=\frac{1}{T}\int_0^{T} dt |\langle h_{1234}\rangle|^2$, where $h_{1234}$ is obtained from third-order perturbation theory, whose detail and its exact form are presented in the Supplemental Material \cite{Supp}. It yields
	
	\begin{widetext}
		\begin{eqnarray}
		\bar{G}_{l_{1,a},l_{4,a}} &=& a_0 +a_1 \langle \mathrm{i}\gamma_{0,1}\gamma_{0,2}\rangle \sin\left[\frac{e}{\hbar}(\Phi_{1,2}-\Phi^{00}_{1,2})\right] +a_2 \langle \mathrm{i}\gamma_{0,3}\gamma_{0,4}\rangle \sin\left[\frac{e}{\hbar}(\Phi_{4,3}-\Phi^{00}_{4,3})\right] \nonumber \\
		&& +a_3 \langle \gamma_{0,1}\gamma_{0,2}\gamma_{0,3}\gamma_{0,4}\rangle \cos\left[\frac{e}{\hbar}(\Phi_{4,3}-\Phi^{00}_{4,3}-\Phi_{1,2}+\Phi^{00}_{1,2})\right]\;, \label{joint}
		\end{eqnarray}
	\end{widetext}
	where $a_0$, $a_1$, $a_2$, $a_3$, $\Phi^{00}_{4,3}$, and $\Phi^{00}_{1,2}$ depend on the system parameters, and we have assumed time independent fluxes $\Phi_{4,3}$ and $\Phi_{1,2}$ for simplicity.
	By tuning the enclosed fluxes $\Phi_{1,2}\approx \Phi_{1,2}^{00}$ and $\Phi_{4,3}\approx \Phi_{4,3}^{00}$, the second and third terms of Eq.~(\ref{joint}) are strongly suppressed, whereas the last term's contribution is maximized. Physically, this corresponds to the following situation: as particles travel from lead $l_{1,a}$ to $l_{4,a}$ through three paths labeled by A: $l_{1,a}\rightarrow l_{4,a}$, B: $l_{1,a}\rightarrow l_{2,a}\rightarrow l_{4,a}$ and C: $l_{1,a}\rightarrow l_{3,a}\rightarrow l_{4,a}$, interference between paths A and B or between paths A and C is always destructive whereas interference between paths B and C revealing four-Majorana parity information is always constructive (see Fig.~\ref{lead2}(b)). Measuring $\bar{G}_{l_{1,a},l_{4,a}}$ then yields only two possible outcomes and corresponds to the measurement of the joint parity $\gamma_{0,1}\gamma_{0,2}\gamma_{0,3}\gamma_{0,4}$, but not $\mathrm{i}\gamma_{0,1}\gamma_{0,2}$ or $\mathrm{i}\gamma_{0,3}\gamma_{0,4}$ individually.
	
	
	As we have developed in detail in the Supplemental Material \cite{Supp}, the {\it forced} measurements of $\gamma_{0,1}\gamma_{0,2}\gamma_{0,3}\gamma_{0,4}$ and any pairs of Majorana modes are sufficient to implement {\it all} Clifford gates. For example, a CNOT gate with the first (second) qubit being the control (target) qubit can be implemented by preparing the third qubit in a $\sigma_z^{(3)}$ eigenstate, which can be accomplished by measuring $\gamma_{0,1}\gamma_{0,2}\gamma_{0,3}\gamma_{0,4}$, then subsequently measuring $\mathrm{i} \gamma_{0,4}\gamma_{\pi,4}$, $\mathrm{i} \gamma_{\pi,2}\gamma_{\pi,4}$, $\mathrm{i}\gamma_{0,3}\gamma_{\pi,2}$, and $\gamma_{0,1}\gamma_{0,2}\gamma_{0,3}\gamma_{0,4}$ in this exact order. {If their parity measurement outcomes do not agree with a prescribed sequence, we only need to perform some additional similar measurements, which are all explicitly worked out in Supplementary Material \cite{Supp}.}
	
	Finally, it is noted that a combination of $T$-gate and Clifford gates is required for quantum universality \cite{magic,magic2,magic3,magic4,magic5}. On the other hand, since the fusion and braiding rules of Majorana modes do not allow the implementation of topologically protected $T$-gate, nontopological dynamical protocols are often employed \cite{magic3,magic4,magic5}. In our system, a geometrical protocol developed by us in Refs.~\cite{braid6,braid7} can in principle be applied between $\gamma_{0,4}$ and $\gamma_{\pi,4}$ to create a magic state \cite{magic,magic2,magic3,magic4,magic5}. As detailed in the Supplemental Material \cite{Supp}, a series of measurements can then be devised to realize a $T$-gate, consuming the magic state in the process.

	\textit{Conclusion.} In view of many recent studies on SOTSCs \cite{HTI7,HTI8,HTI9,HTI10,HTI11,HTI12} and Floquet higher-order topological phases \cite{FHTI1,FHTI2,FHTI3,FHTI4}, we take a major step forward by revealing their great potential for measurement-based quantum computation. We have presented a minimal Floquet SOTSC model with four MZMs and four MPMs, all of which are utilized to encode three qubits that can be systematically measured via coupling the system with external leads and observing the conductance between two of the leads.
	
	As compared with existing measurement-based protocols proposed using first-order topological materials \cite{mea3,mr1,mr2,mr3,mr4,mr5}, our proposal in principle offers two advantages. First, since a single Floquet SOTSC system can already host eight Majorana modes, it naturally forms a Majorana Cooper-pair Box (MCB), which represents a building block for Majorana-based surface codes \cite{mr1,mr5} and practical qubit architectures. Not only such Floquet SOTSC-based MCBs can encode more qubits as compared with other existing MCB proposals, the use of a single system as opposed to an array of nanowires naturally allows all Majorana modes to experience a common charging Hamiltonian and prevents mutual capacitive coupling that may arise between nanowires in an MCB \cite{mr1,mr2}. Second, in the literature complex designs are often required to host multiple Majorana modes by first-order topological materials and as such four-Majorana parity measurements are often difficult to perform. Indeed previously entangling gates were implemented dynamically \cite{mr3,mr4}, which may require a very precise control and knowledge over all system parameters. By contrast, the simplicity of our protocol allows two- and four-Majorana parity measurements to be implemented in a similar manner.
	
	{As a final note,  the general idea of Majorana parity measurements presented above, when cast in a reduced form,
		can equally apply to static SOTSC systems.  As a first step towards realizing the above proposal,
		initial experimental studies can indeed consider static SOTSCs to first confirm the corner-state based measurement schemes developed in this work.
		However, since such static SOTSCs can only host four MZMs at their corners, the implementation of quantum gate operations proposed in this work may not be possible in a single system; a collection of static SOTSC-based MCBs would be required to perform a computational task. This understanding further highlights the advantage of Floquet SOTSCs for measurement-based quantum computation with minimal spacetime overhead, thus motivating future theoretical and experimental studies of Floquet topological quantum computing.}
	
	\begin{acknowledgements}
		{\bf Acknowledgement}: J.G. acknowledges fund support by the Singapore NRF grant No. NRFNRFI2017-
		04 (WBS No. R-144-000-378-281).
	\end{acknowledgements}

\onecolumngrid
\appendix

\vspace{1cm}

\begin{center} {\bf Supplemental Material} \end{center}

This Supplemental Material contains four sections. In Sec.~A, We elaborate the additional properties of the minimal Floquet SOTSC model (Eq.~(1)) presented in the main text, which include its symmetries and corner modes. In Sec.~B, we generalize the notion of Majorana modes to Sambe space and derive the typical form of MZMs and MPMS in harmonically driven topological superconductors. In Sec.~C, we develop a Floquet perturbation theory and apply it to derive the effective Hamiltonian describing a leads-Floquet SOTSC system when two and four leads couplings are switched on. In Sec.~D, we present the implementation of quantum gate operations by a series of measurements.

\section*{Section A: Additional properties of the minimal Floquet SOTSC model}

Since Eq.~(1) in the main text is time periodic, Floquet theory \cite{Flo1,Flo2} can be applied to analyse its properties. In the first quantized language, we may define Floquet Bogoliubov-de Gennes (BdG) Hamiltonian $[\mathcal{H}]_{nm}=h_{\rm BdG}^{(n-m)}+n \hbar \omega \delta_{n,m}$ in an enlarged Hilbert space (Sambe space), where $h_{\rm BdG}^{(j)}$ is the $j$th-order Fourier component of $h_{\rm BdG}(t)$ satisfying

\begin{equation}
H(t)=\frac{1}{2} \left(c_{1,1}^\dagger \cdots c_{2N_x,2N_y} \right) h_{\rm BdG}(t) \left(c_{1,1}^\dagger \cdots c_{2N_x,2N_y} \right)^\dagger \;. \label{fq}
\end{equation}

\begin{figure}
	\begin{center}
		\includegraphics[scale=0.5]{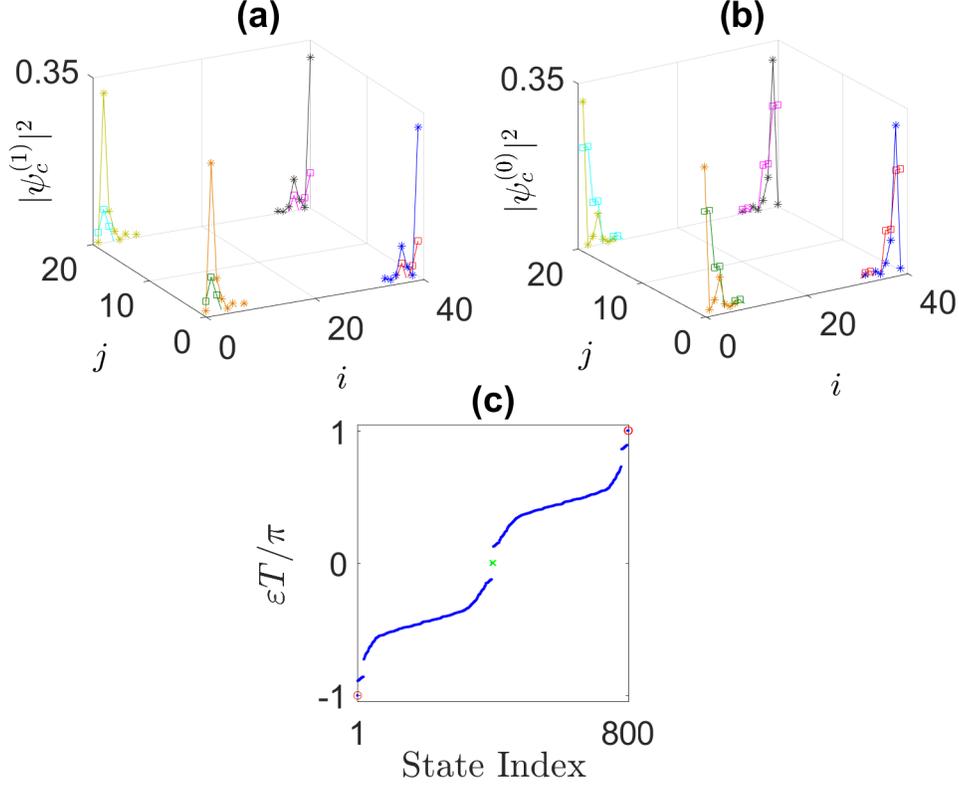}
	\end{center}
	\caption{Probability distributions of the first and zeroth order Fourier components of MZMs (squares) and MPMs (asterisks) are shown in panel (a) and (b) respectively. Panel (c) shows the quasienergy level distribution in the Floquet Brillouin zone, where MZMs (MPMs) are highlighted in green (red). System parameters are chosen as $J_x T/\hbar=\pi/2+0.3$, $\Delta_x T/\hbar=\pi/2-0.2$, $J_y T/\hbar=0.15$, $\delta J_y T/\hbar=0.05$, $\Delta_y T/\hbar=0.55$, $\delta_y T/\hbar= 0.45$, $\mu_0 T/\hbar=\pi/2+0.12$, $\delta \mu_0 T/\hbar=0.02$, $\mu_1 T/\hbar=4$, and $\delta \mu_1 T/\hbar=0$. }
	\label{result1}
\end{figure}

\n Diagonalizing $\mathcal{H}$ gives the set of Floquet eigenstates and their associated excitation quasienergies, i.e., $\left\lbrace|\varepsilon\rangle ;  \varepsilon \right\rbrace$. Due to the time periodicity, the quasienergies $\varepsilon$ and $\varepsilon+\hbar \omega$ describe the same physics and it is thus sufficient to restrict $\varepsilon$ in the Floquet Brillouin zone $\left(-\frac{\hbar \omega}{2}, \frac{\hbar \omega}{2}\right]$. MZMs and MPMs are characterized by the existence of degenerate $\varepsilon=0$ and $\frac{\hbar \omega}{2}$ solutions under OBCs which are localized near the corner of the system, as depicted in Fig.~\ref{result1}(c). By explicitly plotting the probability distributions of these $\varepsilon=0$ and $\frac{\hbar \omega}{2}$ solutions, i.e., Figs.~\ref{result1}(a) and (b), we find that they are indeed sharply localized at one corner of the system, thus confirming the fact that these MZMs and MPMs are in fact Majorana corner modes. There, $|\psi_c^{(0)}|^2$ and $|\psi_c^{(1)}|^2$ describe the probability distributions associated with zeroth and first order Fourier components of the $\varepsilon=0$ and $\frac{\hbar \omega}{2}$ eigenstates, where a full eigenstate is written as $|\Psi_c(t)\rangle = e^{-\mathrm{i}n\omega t/2}|\psi_c(t)\rangle$, $|\psi_c(t)\rangle = \sum_j e^{\mathrm{i} j\omega t}|\psi_c^{(j)}\rangle$, and $n=0$ ($n=1$) for $\varepsilon=0$ ($\varepsilon=\frac{\hbar \omega}{2}$) eigenstate.

Under periodic boundary conditions, Eq.~(\ref{fq}) can be recast in momentum space as

\begin{eqnarray}
h_{\rm BdG,\mathbf{k}}(t) &=& \left[-\left(J_{y,-}+ J_{y,+}\cos(k_y) \right)\sigma_x - J_{y,+} \sin(k_y) \sigma_y -2 J_x \cos(k_x) \right. \nonumber \\
&& \left. +\left(\mu_0+\mu_1 \cos(\omega t)\right)+ \left(\delta\mu_0+\delta \mu_1 \cos(\omega t)\right)\sigma_z \right]\eta_z \nonumber \\
&& +\left[\left(\Delta_{y,-}- \Delta_{y,+}\cos(k_y) \right)\sigma_y +\Delta_{y,+}\sin(k_y)\sigma_x \right] \eta_x + 2\Delta_x \sin(k_x) \eta_y \;, \label{modp}
\end{eqnarray}
where $\sigma$ and $\eta$ are Pauli matrices acting in the sublattice and Nambu space respectively, $J_{y,\pm}= J_y \pm \delta J$, $\Delta_{y,\pm}=\Delta_y \pm \delta_y$, and the lattice spacing is set to unity. In particular, particle-hole symmetry satisfying $\mathcal{P}^{-1} h_{BdG,\mathbf{k}}\mathcal{P}=-h_{BdG,-\mathbf{k}}$ with $\mathcal{P}=\eta_x \mathcal{K}$ is the only symmetry of Eq.~(\ref{modp}), which places the system in the D class of the Altland-Zirnbauer (AZ) classification scheme \cite{AZ}. It is known that 2D first-order topological insulators/superconductors in the D class is characterized by a $Z$ topological invariant (the Chern number), which manifests itself as the number of chiral edge states under OBC.

If $J_{y,\pm}=\delta \mu_0=\delta \mu_1=0$ and $\Delta_{y,-}=\Delta_{y,+}$, the edge of the system parallel to the $y$-direction is equivalent to two copies of that of chiral topological superconductors with opposite Chern numbers, which can be easily verified by diagonalizing $\sigma$ Pauli matrices. By tuning away $\Delta_{y,-}\neq\Delta_{y,+}$, these chiral edge states will then acquire a gap. In this case, additional chiral $\mathcal{C}$, time-reversal $\mathcal{T}$, and inversion $\mathcal{I}$ symmetries also emerge as $\mathcal{C}=\sigma_z \eta_x $, $\mathcal{T}=\sigma_z \mathcal{K}$, and $\mathcal{I}=\sigma_x\eta_z$, which satisfy $\mathcal{C}^{-1}h_{BdG,\mathbf{k}} \mathcal{C}=-h_{BdG,\mathbf{k}}$, $\mathcal{T}^{-1}h_{BdG,\mathbf{k}} \mathcal{T}=h_{BdG,-\mathbf{k}}$, and $\mathcal{I}^{-1}h_{BdG,\mathbf{k}} \mathcal{I}=h_{BdG,-\mathbf{k}}$. In particular, due to the emergence chiral symmetry, two chiral winding numbers $\nu_{0(\pi)}$ can be defined in the spirit of Refs.~\cite{cref,FHTI2}, which determine the presence of Majorana corner states at quasienergy zero $(\pi/T)$ when another edge in the $y$-direction is introduced. Interestingly, although the presence of nonzero $J_{y,\pm},\delta \mu_0$, and $\delta \mu_1$ will break all these additional symmetries, any Majorana corner states in the system remain intact even at moderate values of $J_{y,\pm},\delta \mu_0$, and $\delta \mu_1$ (see again Fig.~\ref{result1}(c)), provided no gap closing occurs. The robustness of these Majorana corner states originates from the remaining particle-hole symmetry in the system, which protects the degeneracy at quasienergy zero or $\pi/T$. Therefore, even at the fundamental level, the minimal Floquet SOTSC model proposed in the main text already represents another form of Floquet second-order topological phases which is solely protected by the particle-hole symmetry and complements our recent discovery of Floquet second-order topological phases protected by the chiral symmetry alone \cite{FHTI2}.

\section*{Section B: MZMs and MPMs in harmonically driven topological superconductors}

Let $H(t)$ be a general time-periodic Hamiltonian with period $T$ and $|\psi_{\varepsilon_0}\rangle =\exp(-\mathrm{i} \frac{\varepsilon_0}{\hbar} t)|R\rangle$ be a reference Floquet eigenstate which satisfies \cite{Flo1,Flo2}

\begin{eqnarray}
\left(H-\mathrm{i}\hbar\frac{\partial}{\partial t} \right) |R\rangle &=& \varepsilon_0 |R\rangle \;, \nonumber \\
\sum_m \left(H^{(n-m)}+m\hbar \omega \delta_{n,m} \right) |R^{(m)}\rangle &=& \varepsilon_0 |R^{(n)}\rangle \;, \nonumber \\
\sum_m \mathcal{H}_{n,m} |R^{(m)}\rangle &=& \varepsilon_0 |R^{(n)}\rangle \;, \label{Sambe}
\end{eqnarray}
where $|R^{(m)}\rangle = \frac{1}{T} \int_0^T dt \; |R\rangle \exp\left(-\mathrm{i} 2\pi m t/T\right)$ and $H^{(m)}=\frac{1}{T} \int_0^T dt \; H \exp\left(-\mathrm{i} 2\pi m t/T\right)$ are the Fourier components of $|R\rangle$ and $H$ respectively. In terms of (time-dependent) second quantized operators $\left\lbrace c_\varepsilon^\dagger\right\rbrace$, another Floquet eigenstate of $H$ can be constructed as $|\psi_{\varepsilon_0+\varepsilon}\rangle = c_\varepsilon^\dagger |\psi_{\varepsilon_0}\rangle$, where $\left[\left(H-\mathrm{i}\hbar\frac{\partial}{\partial t} \right), \bar{c}_\varepsilon^\dagger\right]=\varepsilon \bar{c}_\varepsilon^\dagger $ and $\bar{c}_\varepsilon^\dagger =c_\varepsilon^\dagger \exp(\mathrm{i} \frac{\varepsilon}{\hbar} t)$, so that

\begin{eqnarray}
\left(H-\mathrm{i}\hbar\frac{\partial}{\partial t} \right) \bar{c}_\varepsilon^\dagger |R\rangle &=& \left[\varepsilon \bar{c}_\varepsilon^\dagger+ \bar{c}_\varepsilon^\dagger \left(H-\mathrm{i}\hbar\frac{\partial}{\partial t} \right)\right]|R\rangle \;, \nonumber \\
&=& \left(\varepsilon+\varepsilon_0\right) \bar{c}_\varepsilon^\dagger |R\rangle \;.
\end{eqnarray}
Similar to its static counterpart, the presence of particle-hole symmetry associates each positive excitation creation operator $c_\varepsilon^\dagger$ with its negative excitation annihilation operator, i.e., $c_{-\varepsilon}=c_\varepsilon^\dagger$. As a result, quasienergy zero excitations become Hermitian and must therefore come in pairs as MZMs (e.g. $\gamma_0^{(1)}$ and $\gamma_0^{(2)}$); { since a valid fermion operator must necessarily be complex, a pair of MZMs are required to form e.g. $c_0^\dagger =\gamma_0^{(1)}+\mathrm{i}\gamma_0^{(2)}$}. Unlike its static counterpart, however, quasienergy is only defined modulo $\hbar \omega$. By defining $(\varepsilon_0-\hbar \omega/2,\varepsilon_0+\hbar \omega/2]$ as the Floquet Brillouin zone, quasienergy $\hbar \omega/2$ excitations will also become Hermitian and appear in pairs as MPMs. {  It thus follows that MZMs and MPMs can both be expanded as
	
	\begin{eqnarray}
	\gamma_0 &=& \bar{\gamma}_0=\sum_{n=-\infty}^{\infty} \gamma_{0,n} e^{\mathrm{i} n\omega t} \;,\nonumber \\
	\gamma_\pi &=& \exp(-\mathrm{i}\omega t/2) \bar{\gamma}_\pi = \sum_{n=-\infty}^\infty \gamma_{\pi,n} e^{\mathrm{i} (2n-1)\omega/2}\;, \label{mm}
	\end{eqnarray}
	where $\gamma_{0,n}=\gamma_{0,-n}^\dagger$, $\gamma_{\pi,n}=\gamma_{\pi,-n+1}^\dagger$ are (possibly complex) superposition of Majorana operators. While $\gamma_0$ and $\gamma_\pi$ are now time periodic { (with period $T$ and $2T$ respectively)}, they are still Hermitian at any instant as expected for them to be called MZM and MPM respectively. 
	
	The Fourier components $\gamma_{0,n}$ and $\gamma_{\pi,n}$ are typically localized near the edge (corner) for first-(second-)order topological superconductors and decay exponentially with $|n|$. For a harmonically driven topological superconductor described by the Hamiltonian $H(t)=h_0+h_1 \cos\omega t$, this fact can be formally established by explicitly constructing $\gamma_{0,n}$ and $\gamma_{\pi,n}$ perturbatively. To this end, we may first recast $h_0$ and $h_1$ as superposition of Majorana operators of the form $\mathrm{i}\gamma_i \gamma_j$ for $i,j=1,\cdots,2N$, where $2N$ denote the total number of Majorana operators. If there exists a Majorana operator $\gamma_0^{(0)}$ such that $[h_0,\gamma_0^{(0)}]=0$, we may start by writing $\gamma_0=\gamma_0^{(0)}+\gamma_0^{(1)}$, where $\gamma_0^{(1)}$ is $\propto \frac{1}{\hbar\omega}$. Next, we evaluate}

\begin{equation}
\left[\left(H-\mathrm{i}\hbar\frac{\partial}{\partial t}\right) , \gamma_0\right] = \left[h_1,\gamma_0^{(0)}\right]\cos\omega t -\mathrm{i} \hbar \frac{\partial}{\partial t} \gamma_0^{(1)} +\left[h_0+h_1\cos\omega t,\gamma_0^{(1)}\right]\;.
\end{equation}

\n We may then set $\gamma_0^{(1)}=-\mathrm{i}\left[h_1,\gamma_0^{(0)}\right]\frac{\sin\omega t}{\hbar \omega}+\gamma_0^{(2)}$, where $\gamma_0^{(2)}$ is now $\propto \frac{1}{(\hbar \omega)^2}$, to obtain

\begin{equation}
\left[\left(H-\mathrm{i}\hbar\frac{\partial}{\partial t}\right) , \gamma_0\right] = -\mathrm{i} \left[h_0+h_1\cos\omega t,\left[h_1,\gamma_0^{(0)}\right]\right] \frac{\sin\omega t}{\hbar \omega} -\mathrm{i} \hbar \frac{\partial}{\partial t} \gamma_0^{(2)}+ \left[h_0+h_1 \cos\omega t,\gamma_0^{(2)}\right] \;,
\end{equation}
with the commutator now becoming at most $\propto \frac{1}{\hbar \omega}$. Continuing this procedure leads to

\begin{eqnarray}
\gamma_0^{(2)}&=&\left[h_0,\left[h_1,\gamma_0^{(0)}\right]\right]\frac{\cos\omega t}{(\hbar \omega)^2}+\left[h_1,\left[h_1,\gamma_0^{(0)}\right]\right]\frac{\cos 2 \omega t}{(2\hbar \omega)^2} +\gamma_0^{(3)} \;, \nonumber \\
\left[\left(H-\mathrm{i}\hbar\frac{\partial}{\partial t}\right) , \gamma_0\right] &=& \frac{1}{(\hbar \omega)^2} \left\lbrace h_0+h_1\cos\omega t,\left[h_0,\left[h_1,\gamma_0^{(0)}\right]\right]\cos\omega t +\left[h_1,\left[h_1,\gamma_0^{(0)}\right]\frac{\cos2\omega t}{4}\right]\right\rbrace \nonumber \\
&& -\mathrm{i} \hbar \frac{\partial}{\partial t} \gamma_0^{(3)}+ \left[h_0+h_1 \cos\omega t,\gamma_0^{(3)}\right] \;.
\end{eqnarray}
Repeating the above step indefinitely will thus lead to the form of MZM presented in Eq.~(\ref{mm}), for which it is clear that $\gamma_{0,n}$ is of the order of at most $\frac{1}{(\hbar \omega)^n}$. Moreover, since $\gamma_0^{(0)}$ is localized near the edge (corner) for first-(second-)order topological superconductors while $h_0$ and $h_1$ are superposition of local Majorana operators, nested commutators involving them (thence $\gamma_0$) will also generally be localized near the edge (corner), thus explaining the localization of $\gamma_0$ as a whole.

In a similar fashion, MPMs can also be constructed perturbatively by starting with $\bar{\gamma}_\pi = \gamma_\pi^{(0)} + \gamma_\pi^{(0)\dagger} e^{\mathrm{i}\omega t} +\gamma_\pi^{(1)}$, where $\gamma_\pi^{(0)}$ and $\gamma_\pi^{(1)}$ are superposition of Majorana operators to be determined with the latter being the order of at most $\frac{1}{\hbar \omega}$. In particular,

\begin{eqnarray}
\left[\left(H-\mathrm{i}\hbar\frac{\partial}{\partial t}\right) , \bar{\gamma}_\pi\right] &=& \left\lbrace\left[h_0,\gamma_\pi^{(0)}\right]+ \left[h_1,\gamma_\pi^{(0)\dagger}\right]/2  \right\rbrace +\left\lbrace\left[h_0,\gamma_\pi^{(0)\dagger}\right]+ \left[h_1,\gamma_\pi^{(0)}\right]/2 +\gamma_\pi^{(0)\dagger} \hbar \omega \right\rbrace e^{\mathrm{i}\omega t} \nonumber \\
&& +\left[h_1,\gamma_\pi^{(0)}\right]/2 \;e^{-\mathrm{i}\omega t}+\left[h_1,\gamma_\pi^{(0)\dagger}\right]/2 \;e^{2\mathrm{i}\omega t}+\left[\left(H-\mathrm{i}\hbar\frac{\partial}{\partial t}\right) , \gamma_\pi^{(1)}\right] \;, \label{Pi}
\end{eqnarray}
where $\gamma_\pi^{(0)}$ satisfies

\begin{equation}
\left[h_0,\gamma_\pi^{(0)}\right]+ \left[h_1,\gamma_\pi^{(0)\dagger}\right]/2 = \frac{\hbar \omega}{2} \gamma_\pi^{(0)} \;, \label{cond}
\end{equation}
so that the first line of Eq.~(\ref{Pi}) equals to $\frac{\hbar \omega}{2} \left(\gamma_\pi^{(0)} + \gamma_\pi^{(0)\dagger} e^{\mathrm{i}\omega t}\right)$. Next, we take

\begin{equation}
\gamma_\pi^{(1)}=\frac{1}{\hbar \omega}\left\lbrace A\left[h_1,\gamma_\pi^{(0)}\right]/2 \;e^{-\mathrm{i}\omega t} +B \left[h_1,\gamma_\pi^{(0)\dagger}\right]/2 \;e^{2\mathrm{i}\omega t}\right\rbrace +\gamma_\pi^{(2)} \;, \label{1st}
\end{equation}
where $A$ and $B$ are constants and $\gamma_\pi^{(2)}$ is of the order at most $\frac{1}{(\hbar \omega)^2}$. The two constants $A=2/3$ and $B=-2/5$ are chosen so that the correction terms (the first two terms in the second line) of Eq.~(\ref{Pi}) and $-\mathrm{i}\hbar \frac{\partial}{\partial t}\gamma_\pi^{(1)}$ combine to $\frac{\hbar \omega}{2}\gamma_\pi^{(1)}$. We thus obtain

\begin{eqnarray}
\gamma_\pi^{(1)} &=& \left[h_1,\gamma_\pi^{(0)}\right]/(3\hbar \omega) \;e^{-\mathrm{i}\omega t}-\left[h_1,\gamma_\pi^{(0)\dagger}\right]/(5\hbar \omega) \;e^{2\mathrm{i}\omega t} +\gamma_\pi^{(2)} \;, \nonumber \\
\left[\left(H-\mathrm{i}\hbar\frac{\partial}{\partial t}\right) , \bar{\gamma}_\pi\right] &=& \frac{\hbar \omega}{2} \left( \gamma_\pi^{(0)} + \gamma_\pi^{(0)\dagger} e^{\mathrm{i}\omega t} + \left[h_1,\gamma_\pi^{(0)}\right]/(3\hbar \omega) \;e^{-\mathrm{i}\omega t}-\left[h_1,\gamma_\pi^{(0)\dagger}\right]/(5\hbar \omega) \;e^{2\mathrm{i}\omega t}\right)  \nonumber \\
&& +\frac{1}{\hbar \omega}\left\lbrace\left[h_0,\left[h_1,\gamma_\pi^{(0)}\right]\right]/3\; e^{-\mathrm{i} \omega t} + \left[h_1,\left[h_1,\gamma_\pi^{(0)}\right]\right]/3 \left(1+e^{-2\mathrm{i} \omega t}\right) \right. \nonumber \\
&& -\left. \left[h_0,\left[h_1,\gamma_\pi^{(0)\dagger}\right]\right]/5\; e^{2\mathrm{i}\omega t} - \left[h_0,\left[h_1,\gamma_\pi^{(0)\dagger}\right]\right]/5 \left(e^{\mathrm{i} \omega t}+e^{3\mathrm{i} \omega t}\right) \right\rbrace \nonumber \\
&&+\left[\left(H-\mathrm{i}\hbar\frac{\partial}{\partial t}\right) , \gamma_\pi^{(2)}\right]\;,
\end{eqnarray}
where the correction term is now at most of the order $\frac{1}{\hbar \omega}$, which can be further improved to $\frac{1}{(\hbar \omega)^2}$ via $\gamma_\pi^{(2)}$. The same procedure can again be repeated indefinitely to find an exact expression for MPMs which take the form of Eq.~(\ref{mm}), where $\gamma_{\pi,n}$ is of at most of the order of $\frac{1}{(\hbar \omega)^{n-1}}$ Moreover, if $\gamma_\pi^{(0)}$ is localized near the edge (corner), the same argument as before implies that $\gamma_\pi$ is also localized near the edge (corner).

\section*{Section C: Effective Hamiltonian of the leads-Floquet SOTSC system}

\subsection{Floquet perturbation theory}

In order to derive the effective Hamiltonian presented in the main text, we first develop a Floquet perturbation theory (see Ref.~\cite{FPTI} for a more rigorous formalism). Let $H^{(0)}(t)$ be an unperturbed time-periodic Hamiltonian which satisfies the eigenvalue equation in the Sambe space,

\begin{equation}
\sum_m \mathcal{H}^{(0)}_{n,m} |\tilde{\varepsilon}_j^{(0),(m)}\rangle = \varepsilon_j^{(0)} |\tilde{\varepsilon}_j^{(0),(n)}\rangle \;,
\end{equation}
where $|\tilde{\varepsilon}_j^{(0),(m)}\rangle = \frac{1}{T} \int_0^T dt |\varepsilon_j^{(0)}\rangle \exp\left(-\mathrm{i} 2\pi m t/T\right)$, $\mathcal{H}^{(0)}_{n,m}=H^{(0),(n-m)}+m\hbar \omega \delta_{n,m}$, $H^{(0),(m)}=\frac{1}{T} \int_0^T dt \; H^{(0)} \exp\left(-\mathrm{i} 2\pi m t/T\right)$, and $j$ label the quasienergy index. Since $\mathcal{H}^{(0)}_{n,m}$ is Hermitian, the orthonormality condition for the eigenstates, i.e., $\langle \tilde{\varepsilon}_j^{(0)}|\tilde{\varepsilon}_k^{(0)}\rangle =\delta_{j,k}$, is guaranteed. Consequently, in the presence of a small time-periodic perturbation $\lambda V(t)$ which has the same period as $H(t)$, standard perturbation theory approach follows in the Sambe space, which leads to

\begin{eqnarray}
\varepsilon_j &=& \sum_{l=0} \lambda^l \varepsilon_j^{(l)}\;, \nonumber \\
\varepsilon_j^{(l\geq 1)} &=&   \langle \tilde{\varepsilon}_j^{(0)} | \mathcal{V} | \tilde{\varepsilon}_j^{(l-1)}\rangle + \mathcal{N}_l \;, \nonumber \\ 
|\tilde{\varepsilon}_j\rangle &=& \sum_{l=0} \lambda^l |\tilde{\varepsilon}_j^{(l)}\rangle \;,\label{cor}
\end{eqnarray}
where $\mathcal{V}_{n,m}=\frac{1}{T} \int_0^T dt V(t) \exp\left(-\mathrm{i} 2\pi (n-m) t/T\right)$ and $\mathcal{N}_l$ comes from the renormalization of $|\tilde{\varepsilon}_j^{(l-1)}\rangle$ (the $(l-1)$th order correction to the eigenstate $|\tilde{\varepsilon}_j^{(0)}$). By further assuming that $\langle \tilde{\varepsilon}_j^{(0)} | \mathcal{V} | \tilde{\varepsilon}_j^{(0)} \rangle=0$, which will be relevant to the cases considered below, the quasienergy correction up to third order can be written as



\begin{equation}
\delta \varepsilon_j = \lambda^2 \sum_{i\neq j} \frac{\langle \tilde{\varepsilon}_j^{(0)} |\mathcal{V} |\tilde{\varepsilon}_i^{(0)}\rangle \langle \tilde{\varepsilon}_i^{(0)} |\mathcal{V} |\tilde{\varepsilon}_j^{(0)}\rangle}{\varepsilon_j^{(0)}-\varepsilon_i^{(0)}} +\lambda^3 \sum_{i,k\neq j} \frac{\langle \tilde{\varepsilon}_j^{(0)} |\mathcal{V} |\tilde{\varepsilon}_i^{(0)}\rangle \langle \tilde{\varepsilon}_i^{(0)} |\mathcal{V} |\tilde{\varepsilon}_k^{(0)}\rangle \langle \tilde{\varepsilon}_k^{(0)} |\mathcal{V} |\tilde{\varepsilon}_j^{(0)}\rangle}{\left(\varepsilon_j^{(0)}-\varepsilon_i^{(0)}\right)\left(\varepsilon_j^{(0)}-\varepsilon_k^{(0)}\right)} \;. \label{2nde}
\end{equation}
In the time representation, the inner product can be written as $\langle \tilde{\varepsilon}_j^{(0)} |\mathcal{V} | \tilde{\varepsilon}_j^{(0)} \rangle =\frac{1}{T} \int dt \langle \varepsilon_j^{(0)} | V |\varepsilon_i^{(0)}\rangle $, and Eq.~(\ref{2nde}) becomes

\begin{eqnarray}
\delta \varepsilon_j &=& \lambda^2 \sum_{i\neq j} \frac{\langle \langle \varepsilon_j^{(0)} | V |\varepsilon_i^{(0)}\rangle \rangle \langle \langle \varepsilon_i^{(0)} | V |\varepsilon_j^{(0)}\rangle \rangle}{\varepsilon_j^{(0)}-\varepsilon_i^{(0)}} \nonumber \\
&& + \lambda^3 \sum_{i,k\neq j} \frac{\langle \langle \varepsilon_j^{(0)} | V |\varepsilon_i^{(0)}\rangle \rangle \langle \langle \varepsilon_i^{(0)} | V |\varepsilon_k^{(0)}\rangle \rangle \langle \langle \varepsilon_k^{(0)} | V |\varepsilon_j^{(0)}\rangle \rangle}{\left(\varepsilon_j^{(0)}-\varepsilon_i^{(0)}\right)\left(\varepsilon_j^{(0)}-\varepsilon_k^{(0)}\right)}  \;, \label{2nde2}
\end{eqnarray}
where we have used the shorthand notation $\langle\langle \cdots \rangle \rangle = \frac{1}{T} \int_0^T dt \langle \cdots \rangle$.

In the following, it is also useful to define the notion of the effective Hamiltonian, which amounts to replacing the perturbation term $V(t)$ by an effective term
\begin{equation}
H_{\rm eff}(t) = \sum_j \delta \varepsilon_j |\varepsilon_j^{(0)}\rangle \langle \varepsilon_j^{(0)} |\;,
\end{equation}
where $j$ only sums over the quasienergy within the first Floquet Brillouin zone, i.e., $\varepsilon_j \in \left(-\pi/T,\pi/T\right]$. It is easy to check that if the original Hamiltonian satisfies $\left(H^{(0)}-\mathrm{i}\hbar\frac{\partial}{\partial t} \right) |\varepsilon_j^{(0)}\rangle = \varepsilon_j^{(0)} |\varepsilon_j^{(0)}\rangle$, the addition of $H_{\rm eff}(t)$ implies $\left(H^{(0)}+H_{\rm eff}(t)-\mathrm{i}\hbar\frac{\partial}{\partial t} \right) |\varepsilon_j^{(0)}\rangle = \left(\varepsilon_j^{(0)}+\delta \varepsilon_j\right) |\varepsilon_j^{(0)}\rangle$, which leads to the same energy correction as that described in Eq.~(\ref{2nde2}).

Strictly speaking, while $\tilde{H}=H^{(0)}+H_{\rm eff}$ and $H=H^{(0)}+\lambda V$ share the same quasienergies, they have different eigenstates since $\tilde{H}$ is described by $|\varepsilon_j^{(0)}\rangle$ by construction, whereas $H$ is described by $|\varepsilon_j\rangle = \sum_j \lambda^j |\varepsilon_j^{(j)}\rangle$. The two Hamiltonians are therefore only equal up to some unitary transformations $H=U^\dagger \tilde{H} U$, where $U=\sum_j |\varepsilon_j^{(0)}\rangle \langle \varepsilon_j |$. { In the discussion below, due to the large gap between quasienergy zero or $\pi/T$ and the rest of the quasienergy values, the physics is mainly governed by states closest to quasienergy zero or $\pi/T$ (depending on the system's initial condition). This corresponds to projecting $H$ onto a subspace spanned by $|0,d_0\rangle$ or $|\pi,d_\pi\rangle$, where $d_0$ and $d_\pi$ denote the possible degeneracy of the quasienergy zero or $\pi/T$ eigenstates respectively. After applying such a projection, $H$ and $\tilde{H}$ give effectively the same description. In this sense, the effective Hamiltonian $H_{\rm eff}$ thus represents a good approximation to the actual perturbation $\lambda V$. }




\subsection{Two leads case}

Following the main text, suppose two leads $i=l_{s,m}$ and $j=l_{s',m'}$ are switched on, so that the full Hamiltonian is described by Eq.~(3) in the main text. Without loss of generality, we will first consider the case $n_i=n_j=0$ (similar results are obtained when $n_i$ and $n_j$ are any even integers). In this case, unperturbed eigenstates with particle number $N$ closest to quasienergy zero and $\pi/T$ can be written as an antisymmetrized direct product between the system and lead individual eigenstates as (respectively) 

\begin{eqnarray}
|N,0,(s_1,s_2)\rangle &=& \frac{1}{2}|N\rangle \otimes |e\rangle_{\pi}^{(ij)}\otimes \left[|o\rangle_0^{(ij)} \otimes \left(|1,0\rangle_L + e^{\mathrm{i} \xi_0^{(1)}} s_1 s_2|0,1\rangle_L\right) + |e\rangle_0^{(ij)} \otimes \left( e^{\mathrm{i} \xi_0^{(2)}} s_1|1,0\rangle_L + e^{\mathrm{i} \xi_0^{(3)}} s_2|0,1\rangle_L\right) \right]\;, \nonumber \\
|N,\pi,(s_1,s_2)\rangle &=& \frac{1}{2} |N\rangle \otimes |o\rangle_{\pi}^{(ij)}\otimes \left[|o\rangle_0^{(ij)} \otimes \left(|1,0\rangle_L + e^{\mathrm{i} \xi_\pi^{(1)}} s_1 s_2|0,1\rangle_L\right) +|e\rangle_0^{(ij)} \otimes \left(e^{\mathrm{i} \xi_\pi^{(2)}} s_1|1,0\rangle_L + e^{\mathrm{i} \xi_\pi^{(3)}} s_2|0,1\rangle_L\right)\right]\;, \nonumber \\
\label{base1}
\end{eqnarray}
where $s_1,s_2=\pm 1$ label the degeneracy within each $|N,0,(s_1,s_2)\rangle$ or $|N,\pi,(s_1,s_2)\rangle$, $\xi_\alpha^{(m)}$ for $\alpha=0,\pi$ and $m=1,2,3$ are to be determined later. The lead eigenstates are denoted by $|i,j\rangle_L$, which correspond to the particle occupation of leads $i$ and $j$, while the system eigenstates are further broken down into $|N\rangle\otimes |a\rangle_\pi^{(ij)} \otimes |b\rangle_0^{(ij)}$, where $|a\rangle_\pi^{(ij)}$ ($|b\rangle_0^{(ij)}$), with $a,b=o,e$, denote the parity eigenstate formed by MPMs (MZMs) localized near leads $i$ and $j$, i.e., $\mathrm{i}\gamma_{\pi,i}\gamma_{\pi,j}|e\rangle_\pi^{(ij)}=|e\rangle_\pi^{(ij)}$ and $\mathrm{i}\gamma_{\pi,i}\gamma_{\pi,j}|o\rangle_\pi^{(ij)}=-|o\rangle_\pi^{(ij)}$ ($\mathrm{i}\gamma_{0,i}\gamma_{0,j}|e\rangle_0^{(ij)}=|e\rangle_0^{(ij)}$ and $\mathrm{i}\gamma_{0,i}\gamma_{0,j}|o\rangle_0^{(ij)}=-|o\rangle_0^{(ij)}$), and $|N\rangle$ denotes the other degrees of freedom of the system at particle number $N$, which also includes the parities of other MZMs and MPMs at the other corners. In particular, the quasienergy difference between $|N,0,(s_1,s_2)\rangle$ and $|N,\pi,(s_1,s_2)\rangle$ originates from the fact that the two MPM parity states $|e\rangle_\pi^{(ij)}$ and $|o\rangle_\pi^{(ij)}$ differ in quasienergy by $\pi/T$. In the same fashion, eigenstates with particle number $N\pm 1$ can also be written as

\begin{eqnarray}
|N+1,0,s\rangle &=& \frac{1}{\sqrt{2}}|N+1\rangle \otimes |e\rangle_{\pi}^{(ij)}\otimes \left[\left( |o\rangle_0^{(ij)} + s|e\rangle_0^{(ij)}\right) \otimes |0,0\rangle_L \right]\;, \nonumber \\
|N+1,\pi,s\rangle &=& \frac{1}{\sqrt{2}}|N+1\rangle \otimes |o\rangle_{\pi}^{(ij)}\otimes \left[\left( |o\rangle_0^{(ij)} + s|e\rangle_0^{(ij)}\right) \otimes |0,0\rangle_L \right]\;, \nonumber \\
|N-1,0,s\rangle &=& \frac{1}{\sqrt{2}}|N-1\rangle \otimes |e\rangle_{\pi}^{(ij)}\otimes \left[\left( |o\rangle_0^{(ij)} + s|e\rangle_0^{(ij)}\right) \otimes |1,1\rangle_L \right]\;, \nonumber \\
|N-1,\pi,s\rangle &=& \frac{1}{\sqrt{2}}|N-1\rangle \otimes |o\rangle_{\pi}^{(ij)}\otimes \left[\left( |o\rangle_0^{(ij)} + s|e\rangle_0^{(ij)}\right) \otimes |1,1\rangle_L \right]\;, 
\label{base2}
\end{eqnarray}
where $s=\pm 1$. Finally, it is emphasized that although the usual tensor product notations are used in Eqs.~(\ref{base1}) and (\ref{base2}), antisymmetrization of all the basis states is implied to preserve the anticommutation relation of the fermion operators, e.g., $d_s^\dagger \gamma_{s'} |N,0,(s_1,s_2)\rangle = - \gamma_{s'} d_s^\dagger |N,0,(s_1,s_2)\rangle$. 

We now identify the perturbation as $\lambda V(t)=\sum_{s=i,j} \lambda_{0,s} V_{0,s}+\lambda_{\pi,s} V_{\pi,s}$ with $\lambda_{a,s} V_{a,s}=\lambda_{a,s}(t) d_s^\dagger  \gamma_{a,s}(t)e^{-\mathrm{i} \phi}+h.c.$ ($a=0,\pi$, $s=i,j$). In particular, it follows that $\lambda_{\pi,s} V_{\pi,s}$ connects $|N,0,(s_1,s_2)\rangle$ and $|N\pm 1 ,\pi, s\rangle$ which differ in quasienergy by $\pi/T$, thus resulting in a large quasienergy penalty factor when perturbation theory is applied due to $(\varepsilon_j^{(0)}-\varepsilon_i^{(0)})$ in the denominator of Eq.~(\ref{2nde2}) and can be ignored. In the following, we focus on deriving only the terms in the effective Hamiltonian that are $\propto d_j^\dagger d_i$ and its conjugate, which correspond to terms containing both $\lambda_{0,i} V_{0,i}$ and $\lambda_{0,j} V_{0,j}$. To this end, following a standard approach in second order degenerate perturbation theory \cite{sakurai}, we first fix the phases $\xi_\alpha^{i}$ for $\alpha=0,\pi$ and $i=1,2,3$ by imposing the condition


\begin{eqnarray}
0 &=& \sum_{s,m=\pm 1} \frac{1}{\varepsilon_m} \left[\langle \langle N,l,(s_3,s_4)| \lambda_{0,i} V_{0,i} | N+m,l,s\rangle \rangle \langle \langle N+m,l,s| \lambda_{0,j} V_{0,j} | N,l,(s_1,s_2)\rangle \rangle \right.   \nonumber \\
&& +\left. \langle \langle N,l,(s_3,s_4)| \lambda_{0,j} V_{0,j} | N+m,l,s\rangle \rangle \langle \langle N+m,l,s| \lambda_{0,i} V_{0,i} | N,l,(s_1,s_2)\rangle \rangle  \right] \nonumber \\
&=& \frac{1}{4}\left(\frac{1}{\varepsilon_+} + \frac{1}{\varepsilon_-} \right) \left[\mathrm{i} \left(\bar{\lambda}_{0,i,0} \bar{\lambda}_{0,j,0}^* s_1 s_2 e^{\mathrm{i} \xi_l^{(1)}}- \bar{\lambda}_{0,i,0}^* \bar{\lambda}_{0,j,0} s_3 s_4 e^{-\mathrm{i} \xi_l^{(1)}}\right)\right. \nonumber \\
&& - \left. \mathrm{i} \left(\bar{\lambda}_{0,i,0} \bar{\lambda}_{0,j,0}^* s_2 s_3 e^{\mathrm{i} (\xi_l^{(3)}-\xi_l^{(2)})} - \bar{\lambda}_{0,i,0}^* \bar{\lambda}_{0,j,0} s_1 s_4 e^{\mathrm{i} (\xi_l^{(2)}-\xi_l^{(3)})}\right)\right] \;, \nonumber \\ \label{conds}
\end{eqnarray} 
where $\bar{\lambda}_{l,i(j),n}=\frac{1}{T}\int_0^T dt \lambda_{l,i(j)}e^{-\mathrm{i} n\pi t/T}$ for $l=0,\pi$, $s_3\neq s_1$, and/or $s_2\neq s_4$. { In order to arrive at the second line of Eq.~(\ref{conds}), we have also used $e^{\pm \mathrm{i} \phi}|N\rangle = |N\pm 1\rangle$, $\gamma_{0,i}|e\rangle_0^{(ij)} = |o\rangle_0^{(ij)}$, $\gamma_{0,i}|o\rangle_0^{(ij)} = |e\rangle_0^{(ij)}$, $\gamma_{0,j}|o\rangle_0^{(ij)} = \mathrm{i} |e\rangle_0^{(ij)}$, $\gamma_{0,j}|e\rangle_0^{(ij)} = -\mathrm{i} |o\rangle_0^{(ij)}$, $\gamma_{\pi,i}|e\rangle_\pi^{(ij)} = e^{\mathrm{i} \omega t/2}|o\rangle_\pi^{(ij)}$, $\gamma_{\pi,i}|o\rangle_\pi^{(ij)} = e^{-\mathrm{i} \omega t/2}|e\rangle_\pi^{(ij)}$, $\gamma_{\pi,j}|o\rangle_\pi^{(ij)} = \mathrm{i} e^{-\mathrm{i} \omega t/2} |e\rangle_\pi^{(ij)}$, $\gamma_{\pi,j}|e\rangle_\pi^{(ij)} = -\mathrm{i} e^{\mathrm{i} \omega t/2} |o\rangle_\pi^{(ij)}$, and the fact that only terms that do not involve $s$ survive the summation. It can then be checked that by further taking $\xi_l^{(1)}=-\pi/2+\mathrm{Arg}(\bar{\lambda}_{0,i,0}^* \bar{\lambda}_{0,j,0})$, $\xi_l^{(2)}=0$, and $\xi_l^{(3)}=\pi/2+\mathrm{Arg}(\bar{\lambda}_{0,i,0}^* \bar{\lambda}_{0,j,0})$, Eq.~(\ref{conds}) is satisfied regardless of $s_3\neq s_1$, $s_2\neq s_4$, or both. Indeed, if either $s_3\neq s_1$ or $s_2\neq s_4$ but not both, the two terms in the second and third lines of Eq.~(\ref{conds}) cancel each other, whereas if both $s_3\neq s_1$ and $s_2\neq s_4$ are satisfied, the two terms within each round bracket of Eq.~(\ref{conds}) cancel each other.} The effective Hamiltonian is then found as

\begin{eqnarray}
H_{\rm eff}^{(00)} &=& \sum_{s_1,s_2,s,m=\pm 1} \sum_{l=0,\pi}  \frac{1}{\varepsilon_m}  \left[ \langle \langle N,l,(s_1,s_2)| \lambda_{0,i} V_{0,i} | N+m,l,s\rangle \rangle \langle \langle N+m,l,s| \lambda_{0,j} V_{0,j} | N,l,(s_1,s_2)\rangle \rangle    \right.  \nonumber \\
&& +\left.  \langle \langle N,l,(s_1,s_2)| \lambda_{0,j} V_{0,j} | N+m,l,s\rangle \rangle \langle \langle N+m,l,s| \lambda_{0,i} V_{0,i} | N,l,(s_1,s_2)\rangle \rangle  \right]   | N,l,(s_1,s_2) \rangle \langle N,l,(s_1,s_2) |  \nonumber \\
&=& \sum_{s_1,s_2=\pm 1} \sum_{l=0,\pi} \left(\frac{1}{\varepsilon_+} + \frac{1}{\varepsilon_-} \right) s_1 s_2 \left|\bar{\lambda}_{0,i,0}^* \bar{\lambda}_{0,j,0}\right| \times | N,l,(s_1,s_2) \rangle \langle N,l,(s_1,s_2) | \nonumber \\
&=&  \left(\frac{1}{\varepsilon_+} + \frac{1}{\varepsilon_-} \right) |N\rangle \langle N | \otimes \left(|e\rangle_{\pi}^{(ij)} \langle e|+ |o\rangle_{\pi}^{(ij)} \langle o|\right)\otimes \left(|e\rangle_{0}^{(ij)} \langle e|- |o\rangle_{0}^{(ij)} \langle o|\right)  \nonumber \\
&& \otimes \left(\mathrm{i}\bar{\lambda}_{0,i,0}^* \bar{\lambda}_{0,j,0} |0,1\rangle_L \langle 1,0 |-\mathrm{i}\bar{\lambda}_{0,i,0} \bar{\lambda}_{0,j,0}^* |1,0\rangle_L \langle 0,1 | \right) \nonumber \\
&=& -\left(\frac{1}{\varepsilon_+} + \frac{1}{\varepsilon_-} \right) \bar{\lambda}_{0,i,0}^* \bar{\lambda}_{0,j,0} \gamma_{0,i} \gamma_{0,j} d_j^\dagger d_i +h.c. \;, \label{pert1}
\end{eqnarray}
where $\varepsilon_\pm = \varepsilon_N -\varepsilon_{N\pm 1}$ is the quasienergy difference between $|N,l,(s_1,s_2)\rangle$ and $|N\pm 1,l,s\rangle$, and we have identified $d_j^\dagger d_i =|0,1\rangle_L \langle 1,0|$, $d_i^\dagger d_j =|1,0\rangle_L \langle 0,1|$, and $\mathrm{i} \gamma_{0,i}\gamma_{0,j} =|e\rangle_0^{(ij)} \langle e | - |0\rangle_0^{(ij)} \langle 0 | $. The second equality above can be quickly obtained by first noting that under the choice of $\xi_\alpha^{i}$ obtained earlier, each term in the second equality of Eq.~(\ref{conds}) is $\propto \frac{1}{4}\left(\frac{1}{\varepsilon_+} + \frac{1}{\varepsilon_-} \right) \left|\bar{\lambda}_{0,i,0}^* \bar{\lambda}_{0,j,0}\right|$. Since the terms inside the bracket in the first equality of Eq.~(\ref{pert1}) are of the same form as the first equality of Eq.~(\ref{conds}), but with $s_1=s_3$ and $s_2=s_4$, the second equality of Eq.~(\ref{pert1}) immediately follows. The third equality then comes from the fact that only terms not involving $s_1$ and/or $s_2$ survive the summation. Physically, $H_{\rm eff}^{(00)}$ describes a process in which a particle in lead $i$ ($j$) enters the system to occupy the nonlocal fermion formed by the two MZMs near lead $i$ and $j$, while another particle from this nonlocal fermion leaves the system through lead $j$ ($i$). 

By now considering lead energies with $n_i=-n_j=1$ (similar results are again obtained when $n_i$ and $n_j$ are any odd integers), occupying any one of the leads costs a quasienergy of $\pm \pi/T$, so that the new basis states at $N$ particles are 

\begin{eqnarray}
|N,0,(s_1,s_2)\rangle &=& \frac{1}{2} |N\rangle \otimes |o\rangle_{\pi}^{(ij)}\otimes \left[ |o\rangle_0^{(ij)} \otimes \left(e^{\mathrm{i} \omega t}|1,0\rangle_L + e^{\mathrm{i} \xi_0^{(1)}} s_1 s_2|0,1\rangle_L\right) \right. \nonumber \\
&& \left. + |e\rangle_0^{(ij)} \otimes \left( e^{\mathrm{i} \omega t} e^{\mathrm{i} \xi_0^{(2)}} s_1|1,0\rangle_L +  e^{\mathrm{i} \xi_0^{(3)}} s_2|0,1\rangle_L\right)\right]\;, \nonumber \\
|N,\pi,(s_1,s_2)\rangle &=& \frac{1}{2}|N\rangle \otimes |e\rangle_{\pi}^{(ij)}\otimes \left[|o\rangle_0^{(ij)} \otimes \left(|1,0\rangle_L + e^{-\mathrm{i} \omega t} e^{\mathrm{i} \xi_\pi^{(1)}} s_1 s_2|0,1\rangle_L\right) \right. \nonumber \\
&& \left. +|e\rangle_0^{(ij)} \otimes \left(e^{\mathrm{i} \xi_\pi^{(2)}} s_1|1,0\rangle_L + e^{-\mathrm{i} \omega t} e^{\mathrm{i} \xi_\pi^{(3)}} s_2|0,1\rangle_L\right)\right]\;,
\label{base3}
\end{eqnarray}
where the factor of $e^{\mathrm{i}\omega t}$ appears due to $|1,0\rangle_L$ and $|0,1\rangle_L$ having a quasienergy difference of $\omega=2\pi/T$. On the other hand, since $|N\pm 1, 0,s\rangle$ and $|N\pm 1, \pi,s\rangle$ are the same as before, $\lambda_{0,s} V_{0,s}$ now connects two states differing in quasienergy by $\pi/T$, which can thus be ignored, while dominating terms containing both $\lambda_{\pi,i} V_{\pi,i}$ and $\lambda_{\pi,j} V_{\pi,j}$ become our main interest for which second order perturbation theory is to be applied. We again start by fixing the phases $\xi_\alpha^{(m)}$ through the condition (for $s_3\neq s_1$ and/or $s_2\neq s_4$)

\begin{eqnarray}
0&=& \sum_{s,m=\pm 1} \frac{1}{\varepsilon_m} \left[\langle \langle N,l,(s_3,s_4)| \lambda_{\pi,i} V_{\pi,i} | N+m,l,s\rangle \rangle \langle \langle N+m,l,s| \lambda_{\pi,j} V_{\pi,j} | N,l,(s_1,s_2)\rangle \rangle \right.\nonumber \\
&& \left. +\langle \langle N,l,(s_3,s_4)| \lambda_{\pi,j} V_{\pi,j} | N+m,l,s\rangle \rangle \langle \langle N+m,l,s| \lambda_{\pi,i} V_{\pi,i} | N,l,(s_1,s_2)\rangle \rangle\right] \nonumber \\
&=& \frac{1}{4}\left(\frac{1}{\varepsilon_+} + \frac{1}{\varepsilon_-} \right) \left[\mathrm{i} \left(\bar{\lambda}_{\pi,i,1} \bar{\lambda}_{\pi,j,1}^* s_1 s_2 e^{\mathrm{i} \xi_l^{(1)}}- \bar{\lambda}_{\pi,i,-1}^* \bar{\lambda}_{\pi,j,-1} s_3 s_4 e^{-\mathrm{i} \xi_l^{(1)}}\right)  \right. \nonumber \\
&& \left. + \mathrm{i} \left(\bar{\lambda}_{\pi,i,1} \bar{\lambda}_{\pi,j,1}^* s_2 s_3 e^{\mathrm{i} (\xi_l^{(3)}-\xi_l^{(2)})} - \bar{\lambda}_{\pi,i,-1}^* \bar{\lambda}_{\pi,j,-1} s_1 s_4 e^{\mathrm{i} (\xi_l^{(2)}-\xi_l^{(3)})}\right)\right]\;, \nonumber \\
\label{conds2}
\end{eqnarray} 
which implies $\xi_l^{(1)}=\xi_l^{(3)}=-\pi/2+\mathrm{Arg}(\bar{\lambda}_{\pi,i,-1}^* \bar{\lambda}_{\pi,j,-1})$ and $\xi_l^{(2)}=0$, again regardless of $s_3\neq s_1$, $s_2\neq s_4$, or both. Second order Floquet perturbation theory then gives the effective Hamiltonian

\begin{eqnarray}
H_{\rm eff}^{(\pi\pi)} &=& \sum_{s_1,s_2,s,m=\pm 1} \sum_{l=0,\pi}  \frac{1}{\varepsilon_m} \left[ \langle \langle N,l,(s_1,s_2)| \lambda_{\pi,i} V_{\pi,i} | N+m,l,s\rangle \rangle \langle \langle N+m,l,s| \lambda_{\pi,j} V_{\pi,j} | N,l,(s_1,s_2)\rangle \rangle    \right.  \nonumber \\
&& +\left.  \langle \langle N,l,(s_1,s_2)| \lambda_{\pi,j} V_{\pi,j} | N+m,l,s\rangle \rangle \langle \langle N+m,l,s| \lambda_{\pi,i} V_{\pi,i} | N,l,(s_1,s_2)\rangle \rangle  \right]   | N,l,(s_1,s_2) \rangle \langle N,l,(s_1,s_2) |  \nonumber \\
&=& \sum_{s_1,s_2=\pm 1} \left(\frac{1}{\varepsilon_+} + \frac{1}{\varepsilon_-} \right) s_1 s_2 \left|\bar{\lambda}_{\pi,i}^* \bar{\lambda}_{\pi,j}\right| \times \left(  | N,0,(s_1,s_2) \rangle \langle N,0,(s_1,s_2) | - | N,\pi,(s_1,s_2) \rangle \langle N,\pi,(s_1,s_2) |\right)\nonumber \\
&=&  \left(\frac{1}{\varepsilon_+} + \frac{1}{\varepsilon_-} \right) |N\rangle \langle N | \otimes \left(|e\rangle_{\pi}^{(ij)} \langle e|- |o\rangle_{\pi}^{(ij)} \langle o|\right) \otimes \left(|e\rangle_{0}^{(ij)} \langle e|+ |o\rangle_{0}^{(ij)} \langle o|\right)  \nonumber \\
&& \otimes \left(\mathrm{i}e^{-\mathrm{i} \omega t}\bar{\lambda}_{\pi,i,-1}^* \bar{\lambda}_{\pi,j,-1} |0,1\rangle_L \langle 1,0 |-\mathrm{i}e^{\mathrm{i} \omega t} \bar{\lambda}_{\pi,i,1} \bar{\lambda}_{\pi,j,1}^* |1,0\rangle_L \langle 0,1 | \right) \nonumber \\
&=& -\left(\frac{1}{\varepsilon_+} + \frac{1}{\varepsilon_-} \right) \bar{\lambda}_{\pi,i,-1}^* \bar{\lambda}_{\pi,j,-1} \gamma_{\pi,i} \gamma_{\pi,j} e^{-\mathrm{i} \omega t} d_j^\dagger d_i +h.c. \;, 
\end{eqnarray}
which again describes a process in which a particle enters the system from lead $i$ ($j$) and leaves through $j$ ($i$), but is now mediated by the nonlocal fermion formed between two MPMs near lead $i$ and $j$.

Finally, we consider lead energies with $n_i=0$, $n_j=-1$ (with similar results obtained for $n_i$ and $n_j$ being even and odd integers respectively). In this case, occupying lead $j$ ($i$) costs a quasienergy of $\pi/T$ ($0$), and the new basis states at $N$ and $N\pm 1$ particles now become

\begin{eqnarray}
|N,0,(s_1,s_2)\rangle &=& \frac{1}{2} |N\rangle \otimes \left[ |e\rangle_{\pi}^{(ij)}\otimes  |o\rangle_0^{(ij)} \otimes |1,0\rangle_L + e^{\mathrm{i} \xi_0^{(1)}} s_1 s_2 |o\rangle_{\pi}^{(ij)}\otimes  |e\rangle_0^{(ij)} \otimes |0,1\rangle_L +e^{\mathrm{i} \xi_0^{(2)}} s_1 |e\rangle_{\pi}^{(ij)}\otimes  |e\rangle_0^{(ij)} \otimes |1,0\rangle_L \right.  \nonumber \\
&& \left.+ e^{\mathrm{i} \xi_0^{(3)}} s_2 |o\rangle_{\pi}^{(ij)}\otimes  |o\rangle_0^{(ij)} \otimes |0,1\rangle_L \right] \;, \nonumber \\
|N,\pi,(s_1,s_2)\rangle &=& \frac{1}{2} |N\rangle \otimes \left[ |o\rangle_{\pi}^{(ij)}\otimes  |e\rangle_0^{(ij)} \otimes |1,0\rangle_L + e^{-\mathrm{i} \omega t} e^{\mathrm{i} \xi_\pi^{(1)}} s_1 s_2 |e\rangle_{\pi}^{(ij)}\otimes  |o\rangle_0^{(ij)} \otimes |0,1\rangle_L + e^{\mathrm{i} \xi_\pi^{(2)}} s_1  |o\rangle_{\pi}^{(ij)}\otimes  |o\rangle_0^{(ij)} \otimes |1,0\rangle_L   \right.  \nonumber \\
&& \left. + e^{-\mathrm{i} \omega t}e^{\mathrm{i} \xi_\pi^{(3)}} s_2 |e\rangle_{\pi}^{(ij)}\otimes  |e\rangle_0^{(ij)} \otimes |0,1\rangle_L  \right] \;, \nonumber \\
|N-1,\pi,s\rangle &=& \frac{1}{\sqrt{2}} |N-1\rangle \otimes |e\rangle_{\pi}^{(ij)}\otimes \left[\left( |o\rangle_0^{(ij)} + s|e\rangle_0^{(ij)}\right) \otimes e^{-\mathrm{i} \omega t} |1,1\rangle_L \right]\;, \nonumber \\
|N-1,0,s\rangle &=& \frac{1}{\sqrt{2}}|N-1\rangle \otimes |o\rangle_{\pi}^{(ij)}\otimes \left[\left( |o\rangle_0^{(ij)} + s|e\rangle_0^{(ij)}\right) \otimes |1,1\rangle_L \right]\;,
\label{base4}
\end{eqnarray}
while $|N+1,0,s\rangle$ and $|N+1,\pi,s\rangle$ are the same as those defined in Eq.~(\ref{base2}). As a result, the second order effective Hamiltonian is now dominated by a process involving both $\lambda_{0,i} V_{0,i}$ and $\lambda_{\pi,j} V_{\pi,j}$, thus mixing MZM and MPM together. As usual, we start by fixing the phases $\xi_\alpha^{(m)}$ through the condition (for $s_3\neq s_1$ and/or $s_2\neq s_4$)


\begin{eqnarray}
0&=& \sum_{s,m=\pm 1} \frac{1}{\varepsilon_m} \left[\langle \langle N,l,(s_3,s_4)| \lambda_{0,i} V_{0,i} | N+m,l,s\rangle \rangle \langle \langle N+m,l,s| \lambda_{\pi,j} V_{\pi,j} | N,l,(s_1,s_2)\rangle \rangle \right. \nonumber \\
&& +\left.\langle \langle N,l,(s_3,s_4)| \lambda_{\pi,j} V_{\pi,j} | N+m,l,s\rangle \rangle \langle \langle N+m,l,s| \lambda_{0,i} V_{0,i} | N,l,(s_1,s_2)\rangle \rangle \right] \nonumber\\
&=& \frac{1}{4}\left(\frac{1}{\varepsilon_+} + \frac{1}{\varepsilon_-} \right) \left[\mathrm{i} \left(\bar{\lambda}_{0,i,0} \bar{\lambda}_{\pi,j,1}^* s_1 s_2 e^{\mathrm{i} \xi_l^{(1)}}- \bar{\lambda}_{0,i,0}^* \bar{\lambda}_{\pi,j,-1} s_3 s_4 e^{-\mathrm{i} \xi_l^{(1)}}\right) \right. \nonumber \\
&& \left. + \mathrm{i} \left(\bar{\lambda}_{0,i,0} \bar{\lambda}_{\pi,j,1}^* s_2 s_3 e^{\mathrm{i} (\xi_l^{(3)}-\xi_l^{(2)})} - \bar{\lambda}_{0,i,0}^* \bar{\lambda}_{\pi,j,-1} s_1 s_4 e^{\mathrm{i} (\xi_l^{(2)}-\xi_l^{(3)})}\right)\right] \;, \nonumber \\
\label{conds3}
\end{eqnarray}
which implies that $\xi_l^{(1)}=\xi_l^{(3)}=-\pi/2+\mathrm{Arg}(\bar{\lambda}_{0,i,0}^* \bar{\lambda}_{\pi,j,-1})$ and $\xi_l^{(2)}=0$, { again regardless of $s_3\neq s_1$, $s_2\neq s_4$, or both}. By applying second order Floquet perturbation theory, we obtain the effective Hamiltonian

\begin{eqnarray}
H_{\rm eff}^{(0\pi)} &=& \sum_{s_1,s_2,s,m=\pm 1} \sum_{l=0,\pi} \frac{1}{\varepsilon_m}  \left[ \langle \langle N,l,(s_1,s_2)| \lambda_{0,i} V_{0,i} | N+m,l,s\rangle \rangle \langle \langle N+m,l,s| \lambda_{\pi,j} V_{\pi,j} | N,l,(s_1,s_2)\rangle \rangle    \right.  \nonumber \\
&& +\left.  \langle \langle N,l,(s_1,s_2)| \lambda_{\pi,j} V_{\pi,j} | N+m,l,s\rangle \rangle \langle \langle N+m,l,s| \lambda_{0,i} V_{0,i} | N,l,(s_1,s_2)\rangle \rangle  \right]   | N,l,(s_1,s_2) \rangle \langle N,l,(s_1,s_2) |  \nonumber \\
&=& \sum_{s_1,s_2=\pm 1} \left(\frac{1}{\varepsilon_+} + \frac{1}{\varepsilon_-} \right) s_1 s_2 \left|\bar{\lambda}_{0,i,0}^* \bar{\lambda}_{\pi,j,-1}\right| \times \left(  | N,0,(s_1,s_2) \rangle \langle N,0,(s_1,s_2) | - | N,\pi,(s_1,s_2) \rangle \langle N,\pi,(s_1,s_2) |\right)\nonumber \\
&=&  \left(\frac{1}{\varepsilon_+} + \frac{1}{\varepsilon_-} \right) |N\rangle \langle N | \otimes \left(e^{-\mathrm{i} \omega t/2}\left[|eo\rangle^{(ij)} \langle oe| + |ee\rangle^{(ij)} \langle oo|\right]- e^{\mathrm{i} \omega t/2}\left[|oe\rangle^{(ij)} \langle eo|+|oo\rangle^{(ij)} \langle ee|\right]\right)  \nonumber \\
&& \otimes \left(\mathrm{i}e^{-\mathrm{i} \omega t/2}\bar{\lambda}_{0,i,0}^* \bar{\lambda}_{\pi,j,-1} |0,1\rangle_L \langle 1,0 |-\mathrm{i}e^{\mathrm{i} \omega t/2}\bar{\lambda}_{0,i,0} \bar{\lambda}_{\pi,j,1}^* |1,0\rangle_L \langle 0,1 | \right) \nonumber \\
&=& -\left(\frac{1}{\varepsilon_+} + \frac{1}{\varepsilon_-} \right) \bar{\lambda}_{0,i,0}^* \bar{\lambda}_{\pi,j,-1} \gamma_{0,i} \gamma_{\pi,j} e^{-\mathrm{i}\omega t/2} d_j^\dagger d_i +h.c. \;, 
\label{pert3}
\end{eqnarray}  
where $|ab\rangle^{(ij)} \langle cd|=|a\rangle_\pi^{(ij)}\langle c| \otimes |b\rangle_0^{(ij)}\langle d|$. It describes a process in which a particle enters the system from lead $i$ ($j$) and leaves through $j$ ($i$), but is now mediated by the nonlocal fermion formed between a MZM near lead $i$ and a MPM near lead $j$.

To summarize, we have thus derived the effective Hamiltonians presented in Eq.~(4) in the main text. We identify the tunneling amplitudes as

\begin{eqnarray}
T_{i,j}^{(00)}(t) &=& \left(\frac{1}{\varepsilon_+} + \frac{1}{\varepsilon_-} \right) \mathrm{i} \bar{\lambda}_{0,i,0}^* \bar{\lambda}_{0,j,0} \;,\nonumber \\
T_{i,j}^{(\pi\pi)}(t) &=& \left(\frac{1}{\varepsilon_+} + \frac{1}{\varepsilon_-} \right) \mathrm{i} \bar{\lambda}_{\pi,i,-1}^* \bar{\lambda}_{\pi,j,-1} e^{-\mathrm{i}\omega t} \;, \nonumber \\
T_{i,j}^{(0\pi)}(t) &=& \left(\frac{1}{\varepsilon_+} + \frac{1}{\varepsilon_-} \right) \mathrm{i} \bar{\lambda}_{0,i,0}^* \bar{\lambda}_{\pi,j,-1} e^{-\mathrm{i}\omega t/2} 
\label{amp}
\end{eqnarray}
for the cases $n_i=n_j=0$, $n_i=-n_j=1$, and $n_i=n_j+1=0$ respectively. Taking $n_i$ and $n_j$ to be other integers lead to similar forms as Eq.~(\ref{amp}) with other Fourier components $\bar{\lambda}_{a,s,n}$ and $\bar{\lambda}_{a',s',n'}^*$ ($a,a'=0,\pi$ and $s,s'=i,j$) replacing those in Eq.~(\ref{amp}) and additional $e^{\mathrm{i} n\omega t}$ factor may also appear. However, given that $\bar{\lambda}_{0,s,0}$ and $\bar{\lambda}_{\pi,s,\pm 1}$ ($s=i,j$) are the most dominating Fourier components of $\lambda_{0,s}$ and $\lambda_{\pi,s}$ respectively, it might be best to keep the lead energies $E_i=\frac{n_i\hbar \omega}{2}$ and $E_j=\frac{n_j\hbar \omega}{2}$ within $n_i,n_j=0,\pm 1$. Finally, we expect that the same form of effective Hamiltonians (Eq.~(4) in the main text) will also be obtained when $n_i$ and $n_j$ slightly deviate from integer values, where the coupling amplitudes $T_{i,j}^{(00)}(t)$, $T_{i,j}^{(\pi\pi)}(t)$, and $T_{i,j}^{(0\pi)}(t)$ may involve different time-average of the coupling constants $\lambda_{a,s}$ ($a=0,\pi$ and $s=i,j$) which may no longer reflect their Fourier components, and $e^{\mathrm{i} \nu \omega t}$ with noninteger $\nu$ will also appear. In this scenario, the basis states at $N$ or $N\pm 1$ particles are no longer perfectly degenerate, and the effective Hamiltonians are to be derived by applying the Floquet version of the Schrieffer-Wolff transformation \cite{SW1,SW2,SW3} to project the full Hilbert space onto subspace spanned by relevant states closest to quasienergy zero or $\pi$.

\subsection{Four leads case}

Following the main text, suppose four leads labeled $l_{1,a}$, $l_{2,a}$, $l_{3,a}$, and $l_{4,a}$ are switched on with all their energies set to $0$. Since we are interested in finding the conductance between leads $l_{1,a}$ and $l_{4,a}$, we focus on deriving terms in the effective Hamiltonian that is $\propto d_{l_{4,a}}^\dagger d_{l_{1,a}}$ and its conjugate. Similar to the two leads case, a second order process may occur in which a particle from lead $l_{1,a}$ enters the system while another particle leaving the system through lead $l_{4,a}$, which is mediated by the nonlocal fermion formed by $\mathrm{i}\gamma_{0,1}\gamma_{0,4}$. This corresponds to

\begin{equation}
H_{\rm eff}^{(2)}=-\bar{\lambda}_{0,l_{4,a},0}\bar{\lambda}_{0,l_{1,a},0}^*\left(\frac{1}{\varepsilon_+}+\frac{1}{\varepsilon_-}\right) \gamma_{0,1}\gamma_{0,4} d_{l_{4,a}}^\dagger d_{l_{1,a}} +h.c. \;.
\end{equation}

In third order, two parity dependent processes are possible, which correspond to a particle moving from lead $l_{1,a}$ towards $l_{4,a}$ and is mediated by either lead $l_{2,a}$ or $l_{3,a}$. Before applying third order perturbation theory, it is important to again construct the appropriate basis states $|N,l,d_N\rangle$ and $|N\pm 1,l,d_{N\pm 1}\rangle $, where $l=0,\pi$, $d_N$ and $d_{N\pm 1}$ label the respective degeneracy. By assuming that there is only one particle occupying any of the four leads, $|N,l,d_{N}\rangle$ is $32$-fold degenerate, $|N+1,l,d_{N+1}\rangle $ is eightfold degenerate, while $|N-1,l,d_{N-1}\rangle $ is $16$-fold degenerate, and writing down their explicit form will be very tedious. Fortunately, as we will see later, all of the third order processes can be written in a form similar to the second order processes considered before, thus allowing us to use the results obtained previously.

\begin{figure}
	\begin{center}
		\includegraphics[scale=0.9]{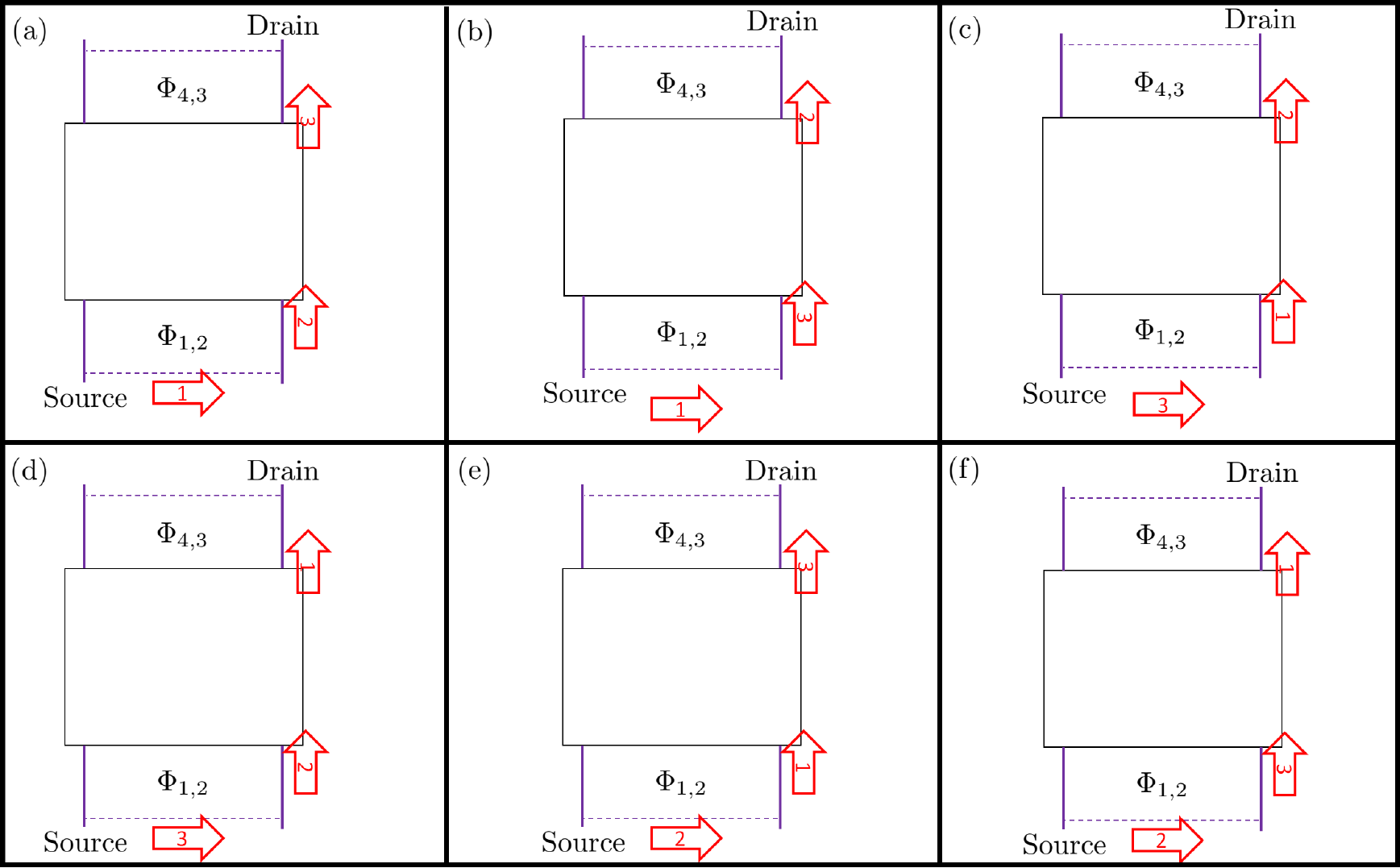}
	\end{center}
	\caption{The six possible subprocesses mediating particle transfer from $l_{1,a}$ to $l_{4,a}$ through $l_{2,a}$, { which are distinguished by different ordering labeled by the numbered arrows $1,2,3$.}}
	\label{lead1}
\end{figure}

We start by considering the process mediated by lead $l_{2,a}$, which can further be broken down into six subprocesses as shown in Fig.~\ref{lead1}. { There, the different subprocesses correspond to different ordering (labeled by the numbered arrows $1,2,3$) for which a particle effectively moves from lead $l_{1,a}$ to $l_{4,a}$. For example, in subprocess b, a particle moves from lead $l_{1,a}$ to $l_{2,a}$, followed by another particle in the system going toward lead $l_{4,a}$, and ended by a particle in lead $l_{2,a}$ entering the system.} To this end, the relevant perturbations are described by $\lambda_{0,l_{2,a}} V_{0,l_{2,a}} = d_{l_{2,a}}^\dagger \lambda_{0,l_{2,a}}(t) \gamma_{0,2}(t)e^{-\mathrm{i} \phi}+h.c.$, $\lambda_{0,l_{4,a}} V_{0,l_{4,a}} = d_{l_{4,a}}^\dagger \lambda_{0,l_{4,a}}(t) \gamma_{0,4}(t)e^{-\mathrm{i} \phi}+h.c.$ , and $\lambda_{(12)} V_{(12)}=\tilde{\lambda}_{l_{1,a},l_{2,a}}^* d_{l_{2,a}}^\dagger d_{l_{1,a}} +h.c.$, where $\tilde{\lambda}_{l_{1,a},l_{2,a}}=\lambda_{l_{1,a},l_{2,a}}e^{\mathrm{i} e\Phi_{1,2}/\hbar}$. It follows that subprocesses described by panel (a)-(e) are zero and will not contribute to the effective Hamiltonian. Consider for example subprocess (a) and (b), which are described by terms such as


\begin{eqnarray}
&&\sum_{d_N'\neq d_N, d_{N\pm 1}}  \langle \langle N,l,d_N| \lambda_{(12)} V_{(12)} | N,l,d_N'\rangle \rangle \left[ \langle \langle N,l,d_N'| \lambda_{0,l_{2,a}} V_{0,l_{2,a}} | N\pm 1,l,d_{N\pm 1}\rangle \rangle  \langle \langle N\pm 1,l,d_{N\pm 1}| \lambda_{0,l_{4,a}} V_{0,l_{4,a}} | N,l,d_N\rangle \rangle \right. \nonumber \\
&& +\left. \langle \langle N,l,d_N'| \lambda_{0,l_{4,a}} V_{0,l_{4,a}} | N\pm 1,l,d_{N\pm 1}\rangle \rangle  \langle \langle N\pm 1,l,d_{N\pm 1}| \lambda_{0,l_{2,a}} V_{0,l_{2,a}} | N,l,d_N\rangle \rangle \right] 
\end{eqnarray}
In particular, the terms inside the square bracket is zero by the choice of basis in the second-order degenerate perturbation theory. Such terms also appear in subprocesses (c) and (d), thus explaining also why they are zero. On the other hand, subprocess (e) is also zero due to the fact that there is only one particle in the leads, e.g. if the particle initially occupies lead $l_{1,a}$, the first step in subprocess (e) is impossible as there is no particle in lead $l_{2,a}$ that can enter the system. This leaves us only with subprocess (f), which leads to

\begin{eqnarray}
H_{\rm eff,1}^{(3)} &=& \frac{1}{\varepsilon_-^2} \sum_{d_N, d_{N-1},d_{N-1}'}\sum_{l=0,\pi} \left\lbrace \langle\langle N,l,d_N| \lambda_{0,l_{2,a}} V_{0,l_{2,a}} | N-1,l,d_{N-1}\rangle \rangle \langle \langle N-1,l,d_{N-1}| \lambda_{(12)} V_{(12)} | N-1,l,d_{N-1}'\rangle \rangle \right. \nonumber \\
&& \times  \left. \langle N-1,l,d_{N-1}'| \lambda_{0,l_{4,a}} V_{0,l_{4,a}} | N,l,d_N\rangle \rangle +c.c.\right\rbrace | N,l,d_N\rangle \langle N,l,d_N | \nonumber \\
&=& \frac{1}{\varepsilon_-^2} \sum_{d_N, d_{N-1}}\sum_{l=0,\pi} \left\lbrace \tilde{\lambda}_{l_{1,a},l_{2,a},0}^* \langle\langle N,l,d_N| d_{l_{1,a}} \lambda_{0,l_{2,a}}^* \gamma_{0,2} e^{\mathrm{i} \phi}  | N-1,l,d_{N-1}\rangle \rangle  \right. \nonumber \\
&& \left. \times \langle \langle N-1,l,d_{N-1}| d_{l_{4,a}}^\dagger \lambda_{0,l_{4,a}} \gamma_{0,4} e^{-\mathrm{i} \phi} | N,l,d_N\rangle \rangle +c.c. \right\rbrace  | N,l,d_N\rangle \langle N,l,d_N | \nonumber \\
&=& -\frac{1}{\varepsilon_-^2} \tilde{\lambda}_{l_{1,a},l_{2,a},0}^* \bar{\lambda}_{0,l_{4,a},0} \bar{\lambda}_{0,l_{2,a},0}^* \gamma_{0,2} \gamma_{0,4} d_{l_{4,a}}^\dagger d_{l_{1,a}} +h.c. \;, \label{3rd1}
\end{eqnarray} 
where $\tilde{\lambda}_{l_{1,a},l_{2,a},n}=\frac{1}{T}\int_0^T dt \tilde{\lambda}_{l_{1,a},l_{2,a}} e^{-\mathrm{i} n\omega t/2}$ and we have also used the fact that $\langle N,l,d_N | d_{l_{2,a}}d_{l_{2,a}}^\dagger = \langle N,l,d_N |$, $\langle N,l,d_N | d_{l_{2,a}}^\dagger d_{l_{2,a}}^\dagger = 0$, and

\begin{eqnarray*}
	\langle \langle N-1,l,d_{N-1}| \lambda_{(12)} V_{(12)} | N-1,l,d_{N-1}'\rangle \rangle |N-1,l,d_{N-1}\rangle \langle N-1,l,d_{N-1}'| &=& \delta_{d_{N-1}',d_{N-1}} \left(\tilde{\lambda}_{l_{1,a},l_{2,a},0}^* d_{l_{2,a}}^\dagger d_{l_{1,a}}+h.c.\right) \\
	&& \times |N-1,l,d_{N-1}\rangle \langle N-1,l,d_{N-1}'| 
\end{eqnarray*}
to get the second line in Eq.~(\ref{3rd1}), after which it reduces to a second-order like expression which allows us to obtain the final result.

\begin{figure}
	\begin{center}
		\includegraphics[scale=0.9]{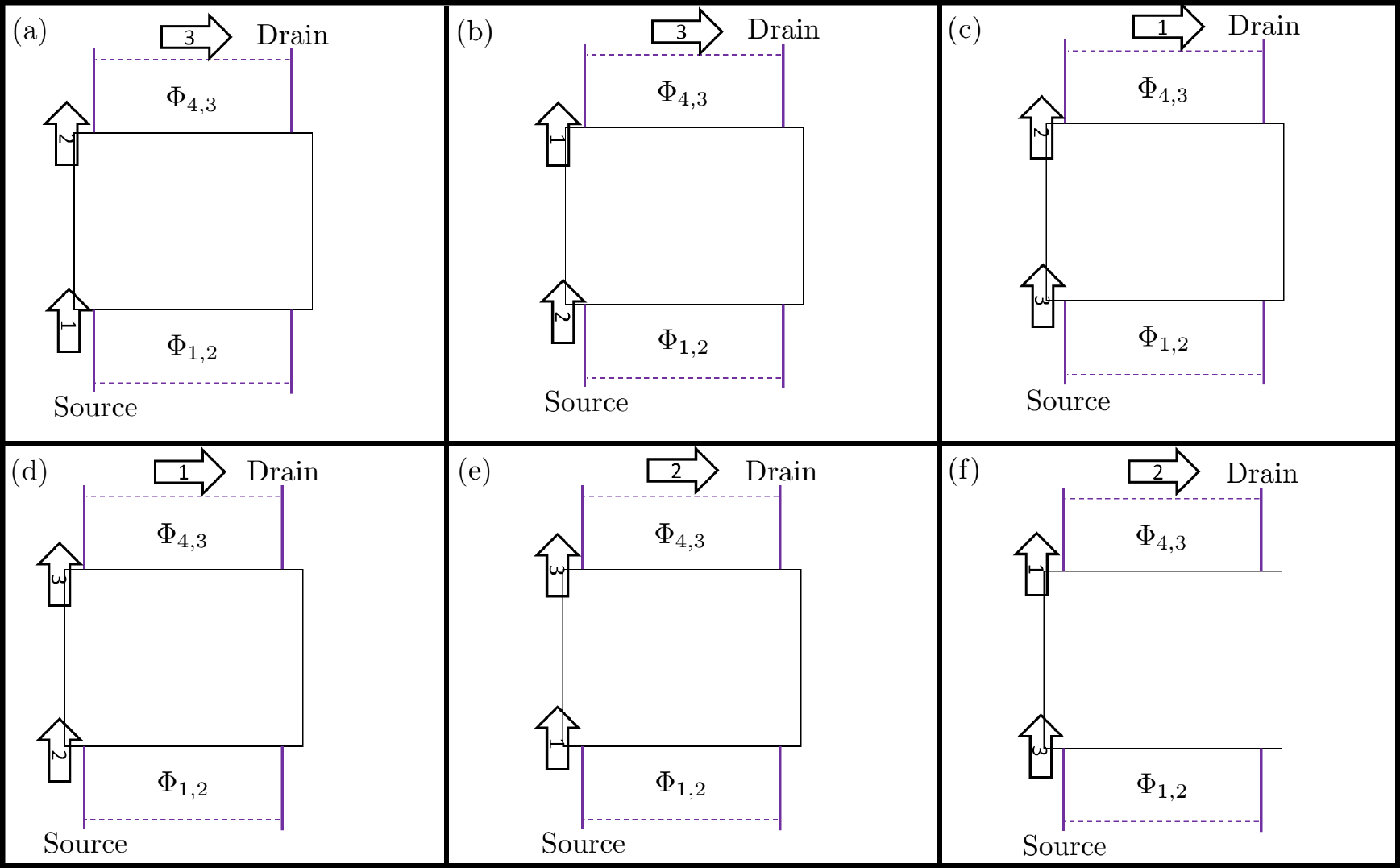}
	\end{center}
	\caption{The six possible subprocesses mediating particle transfer from $l_{1,a}$ to $l_{4,a}$ through $l_{3,a}$, { which are distinguished by different ordering labeled by the numbered arrows $1,2,3$.}}
	\label{lead2}
\end{figure}

In a similar fashion, the process mediated by lead $l_{3,a}$ can also be broken down into six subprocesses, { each corresponding to different ordering for which a particle moves from $l_{1,a}$ to $l_{4,a}$, as shown in Fig.~\ref{lead2}. Among these subprocesses, only that described by panel (f) contributes to the effective Hamiltonian as}

\begin{eqnarray}
H_{\rm eff,2}^{(3)} &=& -\frac{1}{\varepsilon_-^2} \tilde{\lambda}_{l_{3,a},l_{4,a},0}^* \bar{\lambda}_{0,l_{3,a},0} \bar{\lambda}_{0,l_{1,a},0}^* \gamma_{0,1} \gamma_{0,3} d_{l_{4,a}}^\dagger d_{l_{1,a}} +h.c. \;, \label{3rd2}
\end{eqnarray}

Up to the third order, we thus have

\begin{eqnarray}
H_{\rm eff}^{(1234)} &=& h_{1234} d_{l_{4,a}}^\dagger d_{l_{1,a}} +h.c. \;, \nonumber \\
h_{1234} &=& -\bar{\lambda}_{0,l_{4,a},0}\bar{\lambda}_{0,l_{1,a},0}^*\left(\frac{1}{\varepsilon_+}+\frac{1}{\varepsilon_-}\right) \gamma_{0,1}\gamma_{0,4} -\frac{1}{\varepsilon_-^2} \left( \tilde{\lambda}_{l_{1,a},l_{2,a},0}^* \bar{\lambda}_{0,l_{4,a},0} \bar{\lambda}_{0,l_{2,a},0}^* \gamma_{0,2} \gamma_{0,4}  + \tilde{\lambda}_{l_{3,a},l_{4,a},0}^* \bar{\lambda}_{0,l_{3,a},0} \bar{\lambda}_{0,l_{1,a},0}^* \gamma_{0,1} \gamma_{0,3}\right)\;. \nonumber \\ \label{third} 
\end{eqnarray}



\section*{Section D: Measurement-only implementation of Clifford gates}


In this section, we present the implementation of Clifford gates by using a series of Majorana measurements whose net effect is equivalent to braiding. This approach can thus also be interpreted as performing braiding without physically moving the Majorana modes, i.e., braiding by teleportation \cite{mea,mea1,mea2,mea3,mea4}. All three qubits in the system will be needed in this approach, with the first two qubits being the logical qubits and the third qubit being an ancilla. { The ancilla qubit, which is prepared in a $\sigma_z^{(3)}$ eigenstate (unless otherwise specified), is necessary to separate different operators in the logical qubit projectors, thus filtering only the intended unitary operators on the logical qubit states.} Without loss of generality, we will also assume that the system is in the even parity sector, i.e., $\gamma_{0,1}\gamma_{0,2}\gamma_{0,3}\gamma_{0,4}\gamma_{\pi,1}\gamma_{\pi,2}\gamma_{\pi,3}\gamma_{\pi,4}=1$. {While there have been other works showing the implementation of Majorana-measurement-based Clifford gates, our protocols involve only one species of four-Majorana measurements, i.e., $\gamma_{0,1}\gamma_{0,2}\gamma_{0,3}\gamma_{0,4}$, typically performed at the end and start of the protocols, while the remaining steps involve only the measurement of two Majorana operators. Moreover, given that the details of how such measurement protocols work have rarely been explained in other papers, we explicitly elaborate below how each measurement protocol gives the intended outcome.}
{ 
	\vspace{0.3cm}
	
	\textit{General idea.} In order to understand how measurement-based gate operations are possible in the first place, consider the implementation of the $Z_i$-($X_i$-)gate on any one of the logical qubits, i.e., $i=1,2$, via measurements only. Given that a measurement of qubit $\sigma_\alpha^{(i)}$ with outcome $s$ corresponds to applying a projector $\Pi_{i,\alpha}^{s}=\frac{1}{\sqrt{2}}(1+s\sigma_\alpha^{(i)})$, where $\alpha=x,y,z$, to the system, measuring $\sigma_z^{(i)}$ ($\sigma_x^{(i)}$) results in the system transforming as $|\psi\rangle \rightarrow (1+s\sigma_z^{(i)})|\psi\rangle$ ($|\psi\rangle \rightarrow (1+s\sigma_x^{(i)})|\psi\rangle$) up to a normalization factor. Note that if we can find a way to extract only the second term in the projector, the desired $Z_i$-($X_i$)-gate is achieved. To this end, we may utilize the obvious identity $\Pi_{i,\alpha}^{s_1}\Pi_{i,\alpha}^{s_2}=\delta_{s_1,s_2}\Pi_{i,\alpha}^{s_2}$ and the anticommutation relation between Pauli matrices to obtain
	
	\begin{eqnarray}
	\Pi_{3,x}^{s_3}\Pi_{i3,\alpha z}^{s_2}\Pi_{3,x}^{s_1} &\propto& (1+s_3\sigma_x^{(i)})(1+s_2\sigma_\alpha^{(i)}\sigma_z^{(3)})(1+s_1\sigma_x^{(i)}) \nonumber \\
	&\propto & \Pi_{3,x}^{s_1} (s_3+s_1) +s_2\sigma_\alpha^{(i)}\sigma_z^{(3)} \Pi_{3,x}^{s_1} (s_3-s_1) \;,
	\end{eqnarray}
	where $\Pi_{i3,\alpha z}^{s_2}=(1+s_2\sigma_\alpha^{(i)}\sigma_z^{(3)})$ is the projector onto $\sigma_\alpha^{(i)}\sigma_z^{(3)}=s_2$ eigenstate. Namely, up to an overall transformation of the ancilla state, either identity or $\sigma_{\alpha}^{(i)}$ operator is applied to the logical qubit states if $s_3=s_1$ or $s_3=-s_1$ respectively. Physically, the three projectors $\Pi_{3,x}^{s_3}$, $\Pi_{i3,\alpha z}^{s_2}$, and $\Pi_{3,x}^{s_1}$ can be implemented by measuring $\sigma_x^{(3)}$, $\sigma_\alpha^{(i)}\sigma_z^{(3)}$, and $\sigma_x^{(3)}$ in this order, where the intended gate is obtained if the first and third $\sigma_x^{(3)}$ measurements yield different outcome. If they yield the same outcome, identity operator (with respect to the logical qubits) is instead obtained, and the same measurements can be repeated indefinitely until the last $\sigma_x^{(3)}$ yields a different outcome, i.e., 
	
	\begin{equation}
	\Pi_{3,x}^{-s_1} \Pi_{i3,\alpha z}^{s_{2,N}}\cdots \Pi_{3,x}^{s_1} \Pi_{i3,\alpha z}^{s_{2,2}}\Pi_{3,x}^{s_1} \Pi_{i3,\alpha z}^{s_{2,1}} \Pi_{3,x}^{s_1} = \Pi_{3,x}^{-s_1} \Pi_{i3,\alpha z}^{s_{2,N}} \Pi_{3,x}^{s_1}\;.
	\end{equation} 
	This procedure, termed \emph{forced measurements} \cite{mea}, allows the desired $Z_i$-($X_i$)-gate to be obtained with $100$\% success rate. Similar projector tricks presented above will also be the basis in devising measurement protocols for other Clifford gates in the following, where additional $Z_i$- and $X_i$-gates implemented above might also be needed as correction operators. The default encoding presented in the main text will be used to express all Pauli matrices in terms of MZMs and MPMs, and the ancilla qubit is to be initialized in the $\sigma_z^{(3)}=\gamma_{0,1}\gamma_{0,2}\gamma_{0,3}\gamma_{0,4}=\gamma_{\pi,1}\gamma_{\pi,2}\gamma_{\pi,3}\gamma_{\pi,4}=s$ eigenstate.
	
	\vspace{0.3cm}

	\textit{Hadamard gate.} Hadamard gates can be written in terms of Pauli matrices or MZMs and MPMs as $\mathcal{H}_1\propto \sigma_z^{(1)}+\sigma_x^{(1)}=\mathrm{i}\gamma_{0,1}\gamma_{0,2}+\mathrm{i}\gamma_{0,1}\gamma_{0,3}$ and $\mathcal{H}_2\propto \sigma_z^{(2)}+\sigma_x^{(2)}=\mathrm{i}\gamma_{\pi,1}\gamma_{\pi,2}+\mathrm{i}\gamma_{\pi,1}\gamma_{\pi,3}$. By using the projector tricks presented above, we find that $\mathcal{H}_{1(2)}$ can be obtained via measurements as
	
	\begin{eqnarray}
	\mathcal{H}_1 &=& P_1^{s_1,s_2,s_3,s_4} \Pi_{3,z}^{s_5} \Pi_{3,x}^{s_4} \Pi_{13,zz}^{s_3} \Pi_{13,xz}^{s_2} \Pi_{3,x}^{s_1}\;, \nonumber \\
	&\propto &  P_1^{s_1,s_2,s_3,s_4} \left(1+s_5\gamma_{0,1}\gamma_{0,2}\gamma_{0,3}\gamma_{0,4}\right)\left(1+s_4\mathrm{i}\gamma_{0,4}\gamma_{\pi,4}\right)\left(1+s_3\mathrm{i}\gamma_{0,3}\gamma_{0,4}\right) \left(1+s_2\mathrm{i}\gamma_{0,2}\gamma_{0,4}\right)  \left(1+s_1\mathrm{i}\gamma_{0,4}\gamma_{\pi,4}\right)\;, \nonumber \\
	\mathcal{H}_2 &=& P_2^{s_1,s_2,s_3,s_4} \Pi_{3,z}^{s_5} \Pi_{3,x}^{s_4} \Pi_{23,zz}^{s_3} \Pi_{23,xz}^{s_2} \Pi_{3,x}^{s_1}\;, \nonumber \\
	&\propto & P_2^{s_1,s_2,s_3,s_4} \left(1+s_5\gamma_{0,1}\gamma_{0,2}\gamma_{0,3}\gamma_{0,4}\right)\left(1+s_4\mathrm{i}\gamma_{0,4}\gamma_{\pi,4}\right)\left(1+s_3\mathrm{i}\gamma_{\pi,3}\gamma_{\pi,4}\right) \left(1+s_2\mathrm{i}\gamma_{\pi,2}\gamma_{\pi,4}\right) \left(1+s_1\mathrm{i}\gamma_{0,4}\gamma_{\pi,4}\right) \;, \nonumber \\
	P_j^{s_1,s_2,s_3,s_4} &=& \begin{cases}
	1 & \text{if $s_2=-s_3$, $s_1=-s_4$} \\
	X_jZ_j\equiv \Pi_{3,z}^{s_7} \Pi_{3,x}^{s_6} \Pi_{j3,xz}^{s_8} \Pi_{3,x}^{-s_6} \Pi_{j3,zz}^{s_7} \Pi_{3,x}^{s_6} & \text{if $s_2=s_3$, $s_1=-s_4$} \\
	X_j \equiv \Pi_{3,z}^{s_7}\Pi_{3,x}^{-s_6} \Pi_{j3,xz}^{s_7} \Pi_{3,x}^{s_6} & \text{if $s_2=s_3$, $s_1=s_4$} \\
	Z_j \equiv \Pi_{3,z}^{s_7} \Pi_{3,x}^{-s_6} \Pi_{j3,zz}^{s_7} \Pi_{3,x}^{s_6} & \text{if $s_2=-s_3$, $s_1=s_4$}
	\end{cases} \;,
	\label{had}
	\end{eqnarray}
	where each projector above can be implemented through either two- or four-Majorana parity measurement discussed in the main text and the correction operators $P_j^{s_1,s_2,s_3,s_4}$ involve either or both $Z_i$- and $X_i$-gates implementable via forced measurements described above. As pointed out in Ref.~\cite{mea4}, $Z_i$- and $X_i$-gates, and thus $P_{1(2)}^{s_1,s_2,s_3,s_4}$, can in principle also be implemented using a classical computer, thus potentially avoiding the necessity for additional measurements.
	
	For completeness, it is also straightforward to verify that Eq.~(\ref{had}) indeed yields Hadamard gates. To this end, one may directly expand

	\begin{eqnarray}
	\left(1+s_4\mathrm{i}\gamma_{0,4}\gamma_{\pi,4}\right)\left(1+s_3\mathrm{i}\gamma_{0,3}\gamma_{0,4}\right) \left(1+s_2\mathrm{i}\gamma_{0,2}\gamma_{0,4}\right) \left(1+s_1\mathrm{i}\gamma_{0,4}\gamma_{\pi,4}\right)&=&\left(s_2\mathrm{i}\gamma_{0,2}\gamma_{0,4}+s_3\mathrm{i}\gamma_{0,3}\gamma_{0,4}\right) \left(1+s_1\mathrm{i}\gamma_{0,4}\gamma_{\pi,4}\right)(s_1-s_4) \nonumber \\
	&& +\left(1+s_2s_3 \gamma_{0,3} \gamma_{0,2}\right) \left(1+s_1\mathrm{i}\gamma_{0,4}\gamma_{\pi,4}\right)(s_1+s_4) \;.\nonumber \\
	\end{eqnarray}
	Next, one uses again the anticommutation relation to move $\left(1+s_5\gamma_{0,1}\gamma_{0,2}\gamma_{0,3}\gamma_{0,4}\right)$ to the right, then uses the fact that the third qubit is in a definite $\sigma_z^{(3)}=s$ eigenstate to replace $\gamma_{0,1}\gamma_{0,2}\gamma_{0,3}\gamma_{0,4}=s$. This implies
	
	\begin{eqnarray}
	\Pi_{3,z}^{s_5} \Pi_{3,x}^{s_4} \Pi_{13,zz}^{s_3} \Pi_{13,xz}^{s_2} \Pi_{3,x}^{s_1} &=& \left[ \left(s_2\mathrm{i}\gamma_{0,2}\gamma_{0,4}+s_3\mathrm{i}\gamma_{0,3}\gamma_{0,4}\right) (s_1-s_4) +\left(1+s_2s_3 \gamma_{0,3} \gamma_{0,2}\right)(s_1+s_4) \right] (1+s_5 s)\nonumber \\
	&& \left[ \left(s_2\mathrm{i}\gamma_{0,2}\gamma_{0,4}+s_3\mathrm{i}\gamma_{0,3}\gamma_{0,4}\right) (s_1-s_4) +\left(1+s_2s_3 \gamma_{0,3} \gamma_{0,2}\right)(s_1+s_4) \right] (1-s_5 s)\mathrm{i}\gamma_{0,4}\gamma_{\pi,4}\;,\nonumber \\ 
	&=& \left[ \left(ss_2\mathrm{i}\gamma_{0,1}\gamma_{0,3}-ss_3\mathrm{i}\gamma_{0,1}\gamma_{0,2}\right) (s_1-s_4) +\left(1+s_2s_3 \gamma_{0,3} \gamma_{0,2}\right)(s_1+s_4) \right] (1+s_5 s)\nonumber \\
	&& \left[ \left(ss_2\mathrm{i}\gamma_{0,1}\gamma_{0,3}-ss_3\mathrm{i}\gamma_{0,1}\gamma_{0,2}\right) (s_1-s_4) +\left(1+s_2s_3 \gamma_{0,3} \gamma_{0,2}\right)(s_1+s_4) \right] (1-s_5 s)\mathrm{i}\gamma_{0,4}\gamma_{\pi,4}\;, \nonumber \\
	\end{eqnarray}
	where we have used $\mathrm{i} \gamma_{0,2}\gamma_{0,4}=\mathrm{i} \gamma_{0,1}\gamma_{0,3}\left(\gamma_{0,1}\gamma_{0,2}\gamma_{0,3}\gamma_{0,4}\right)$, $\mathrm{i} \gamma_{0,3}\gamma_{0,4}=-\mathrm{i} \gamma_{0,1}\gamma_{0,2}\left(\gamma_{0,1}\gamma_{0,2}\gamma_{0,3}\gamma_{0,4}\right)$, and $\gamma_{0,1}\gamma_{0,2}\gamma_{0,3}\gamma_{0,4}=s$ in the second equality. By noticing that $\mathrm{i}\gamma_{0,4}\gamma_{\pi,4}=\sigma_x^{(3)}$ only affects the ancilla qubit, it follows that Hadamard gate $\mathcal{H}_1$ is automatically attained if $s_1=-s_4$ and $s_2=-s_3$. If $s_1=-s_4$ and $s_2=s_3$, the relative minus sign between $\mathrm{i}\gamma_{0,1}\gamma_{0,3}$ and $\mathrm{i}\gamma_{0,1}\gamma_{0,2}$ can be rectified by applying $X_1Z_1$. Similarly, if $s_1=s_4$, additional $X_1$ or $Z_1$ gate can be applied respectively for $s_2=s_3$ or $s_2=-s_3$. The measurement outcome $s_5$ does not affect the logical qubit and will simply reinitialize the ancilla qubit in the $\sigma_z^{(3)}=s_5$ eigenstate. The same steps can be applied to verify $\mathcal{H}_2$.
	
	
}

\vspace{0.3cm}

\textit{Phase gate.} Phase gates, which are described by the unitaries $\mathcal{P}_1\propto 1+\gamma_{0,1}\gamma_{0,2}$ and $\mathcal{P}_2\propto 1+\gamma_{\pi,1}\gamma_{\pi,2}$, can also be obtained as a series of measurements

\begin{eqnarray}
\mathcal{P}_1 &=& P_1^{s_1,s_2} \Pi_{3,z}^{s_3} \Pi_{13,zy}^{s_2} \Pi_{3,x}^{s_1} \nonumber \\
&\propto & P_1^{s_1,s_2} \left(1+s_3\gamma_{0,1}\gamma_{0,2}\gamma_{0,3}\gamma_{0,4}\right)\times \left(1-s_2\mathrm{i}\gamma_{0,3}\gamma_{\pi,4}\right)\times \left(1+s_1\mathrm{i}\gamma_{0,4}\gamma_{\pi,4}\right) \;, \nonumber \\
\mathcal{P}_2 &=& P_2^{s_1,s_2}\Pi_{3,z}^{s_3} \Pi_{23,zy}^{s_2} \Pi_{3,x}^{s_1} \nonumber \\
&\propto & P_2^{s_1,s_2}\left(1+s_3\gamma_{0,1}\gamma_{0,2}\gamma_{0,3}\gamma_{0,4}\right)\times \left(1+s_2\mathrm{i}\gamma_{\pi,3}\gamma_{0,4}\right)\times \left(1+s_1\mathrm{i}\gamma_{0,4}\gamma_{\pi,4}\right) \;, \nonumber \\
P_j^{s_1,s_2} &=& \begin{cases}
1 & \text{if $s_1=ss_2$} \\
Z_j \equiv \Pi_{3,z}^{s_6} \Pi_{3,x}^{-s_4} \Pi_{j3,zz}^{s_5} \Pi_{3,x}^{s_4} & \text{if $s_1=-ss_2$}
\end{cases} \;,
\end{eqnarray}
where the ancilla qubit is again initialized in the $\sigma_z^{(3)}=s$ eigenstate. The above can be verified by expanding the product $\left(1-s_2\mathrm{i}\gamma_{0,3}\gamma_{\pi,4}\right)\times \left(1+s_1\mathrm{i}\gamma_{0,4}\gamma_{\pi,4}\right)$ ($\left(1+s_2\mathrm{i}\gamma_{\pi,3}\gamma_{0,4}\right)\times \left(1+s_1\mathrm{i}\gamma_{0,4}\gamma_{\pi,4}\right)$), then moving $\gamma_{0,1}\gamma_{0,2}\gamma_{0,3}\gamma_{0,4}$ to the right and replacing it with $s$. This yields

\begin{eqnarray}
\Pi_{3,z}^{s_3} \Pi_{13,zy}^{s_2} \Pi_{3,x}^{s_1} &\propto& \left(1+s_1s_2 \gamma_{0,3}\gamma_{0,4}\right)\times (1+s_3 s)+ \left(s_1\mathrm{i} \gamma_{0,4}\gamma_{\pi,4}-s_2\mathrm{i} \gamma_{0,3}\gamma_{\pi,4}\right)\times (1-s_3 s)\nonumber \\
&=& \left(1-ss_1s_2 \gamma_{0,1}\gamma_{0,2}\right)\times (1+s_3 s)+ \left(s_1-s s_2 \gamma_{0,1}\gamma_{0,2}\right)\times (1-s_3 s)\times \mathrm{i} \gamma_{0,4}\gamma_{\pi,4}\;, \nonumber \\
\Pi_{3,z}^{s_3} \Pi_{23,zy}^{s_2} \Pi_{3,x}^{s_1} &\propto& \left(1+s_1s_2 \gamma_{\pi,3}\gamma_{\pi,4}\right)\times (1+s_3 s)+ \left(s_1\mathrm{i} \gamma_{0,4}\gamma_{\pi,4}+s_2\mathrm{i} \gamma_{\pi,3}\gamma_{0,4}\right)\times (1-s_3 s)\nonumber \\
&=& \left(1-s s_1s_2 \gamma_{\pi,1}\gamma_{\pi,2}\right)\times (1+s_3 s)+ \left(s_1-s s_2\gamma_{\pi,1}\gamma_{\pi,2} \right)\times (1-s_3 s)\times \mathrm{i} \gamma_{0,4}\gamma_{\pi,4}\;.
\end{eqnarray}
In particular, phase gates $\mathcal{P}_{1(2)}$ are automatically attained if $s_1=ss_2$. If $s_1=-ss_2$, one instead obtains the inverse of phase gate, and additional $Z$-gate can be applied via forced measurements that yield $\Pi_{3,z}^{s_6} \Pi_{3,x}^{-s_4} \Pi_{j3,zz}^{s_5} \Pi_{3,x}^{s_4}\propto \sigma_z^{(j)}$.

\vspace{0.3cm}

{ 
	\textit{CNOT gate.} Without loss of generality, consider a CNOT gate in which the first and second qubits being the control and target qubits respectively, which can be represented by the unitary $U_1(X_2)\propto 1+\mathrm{i} \gamma_{0,1}\gamma_{0,2}+\mathrm{i} \gamma_{\pi,1}\gamma_{\pi,3}+ \gamma_{0,1}\gamma_{0,2}\gamma_{\pi,1}\gamma_{\pi,3}$ and realized as a series of measurements
	
	\begin{eqnarray}
	U_1(X_2) &=& P^{s_1,s_2,s_3} \Pi_{3,z}^{s_4} \Pi_{123,zxx}^{s_3} \Pi_{23,xz}^{s_2} \Pi_{3,x}^{s_1} \nonumber \\
	&\propto & P^{s_1,s_2,s_3} \left(1+s_4\gamma_{0,1}\gamma_{0,2}\gamma_{0,3}\gamma_{0,4}\right)\times\left(1+s_3\mathrm{i}\gamma_{0,3}\gamma_{\pi,2}\right)\times\left(1+s_2\mathrm{i}\gamma_{\pi,2}\gamma_{\pi,4}\right)\times\left(1+s_1\mathrm{i}\gamma_{0,4}\gamma_{\pi,4}\right) \;, \nonumber \\
	P^{s_1,s_2,s_3} &=& \begin{cases}
	1 & \text{if $s_1s_3=s s_2=1$} \\
	X_2 \equiv \Pi_{3,z}^{s_7} \Pi_{3,x}^{-s_5} \Pi_{23,xz}^{s_6} \Pi_{3,x}^{s_5} & \text{if $s_1s_3=-s s_2=-1$} \\
	Z_1 \equiv \Pi_{3,z}^{s_7} \Pi_{3,x}^{-s_5} \Pi_{13,zz}^{s_6} \Pi_{3,x}^{s_5} & \text{if $s_1s_3=s s_2=-1$} \\
	Z_1X_2 \equiv \Pi_{3,z}^{s_8} \Pi_{3,x}^{s_5} \Pi_{23,xz}^{s_7} \Pi_{3,x}^{-s_5} \Pi_{13,zz}^{s_6} \Pi_{3,x}^{s_5} & \text{if $s_1s_3=-s s_2=1$}
	\end{cases} \;,
	\end{eqnarray}
	where the ancilla qubit is in the $\sigma_z^{(3)}=s$ eigenstate. The above is verified by expanding $\left(1+s_3\mathrm{i}\gamma_{0,3}\gamma_{\pi,2}\right)\times\left(1+s_2\mathrm{i}\gamma_{\pi,2}\gamma_{\pi,4}\right)\times\left(1+s_1\mathrm{i}\gamma_{0,4}\gamma_{\pi,4}\right)$, then moving $\gamma_{0,1}\gamma_{0,2}\gamma_{0,3}\gamma_{0,4}$ to the right and replacing it by $s$,
	
	\begin{eqnarray}
	\Pi_{3,z}^{s_4} \Pi_{13,zx}^{s_3} \Pi_{23,xz}^{s_2} \Pi_{3,x}^{s_1} &\propto& \left(1+s_2 \mathrm{i} \gamma_{\pi,2}\gamma_{\pi,4} -s_1 s_3  \gamma_{0,3}\gamma_{\pi,2}\gamma_{0,4}\gamma_{\pi,4}+s_1 s_2 s_3 \mathrm{i}\gamma_{0,3}\gamma_{0,4}\right)\times (1+s_4 s) \nonumber \\
	&& + \left(s_1 \mathrm{i} \gamma_{0,4}\gamma_{\pi,4}+s_3 \mathrm{i} \gamma_{0,3}\gamma_{\pi,2}+s_1s_2\gamma_{\pi,2}\gamma_{0,4} -s_2 s_3 \gamma_{0,3}\gamma_{\pi,4}\right)\times (1-s_4 s) \nonumber \\
	&=&\left(1+ss_2 \mathrm{i} \gamma_{\pi,1}\gamma_{\pi,3} +s_1 s_3 \gamma_{0,1}\gamma_{0,2}\gamma_{\pi,1}\gamma_{\pi,3}+ss_1 s_2 s_3 \mathrm{i} \gamma_{0,1}\gamma_{0,2}\right)\times (1+s_4 s) \nonumber \\
	&& + \left(s_1 +s_3\gamma_{0,1}\gamma_{0,2}\gamma_{\pi,1}\gamma_{\pi,3}+ss_1s_2\mathrm{i}\gamma_{\pi,1}\gamma_{\pi,3} +ss_2 s_3 \mathrm{i} \gamma_{0,1}\gamma_{0,2}\right)\times (1-s_4 s)\times \mathrm{i} \gamma_{0,4}\gamma_{\pi,4}
	\end{eqnarray}
	It follows that a CNOT gate is automatically achieved if $s_1 s_3=s s_2=1$. If $s_1 s_3=ss_2=-1$, $s_1 s_3=-ss_2=-1$, or $s_1 s_3=-ss_2=1$, additional $Z_1$-gate, $X_2$-gate, or $Z_1X_2$-gate can be applied respectively via forced measurements.}

\vspace{0.5cm}

Note that at the end of each step above, the ancilla qubit remains in a $\sigma_z^{(3)}$ eigenstate, while the first two (logical) qubits transform according to the intended gate operation. This allows the procedure above to be repeated multiple times to further manipulate the logical qubits in carrying out various quantum computational tasks. Moreover, by preparing the ancilla qubit in the magic state $|M\rangle=\exp\left(-\pi/8\right)|0\rangle_3 + \exp\left(\pi/8\right)|1\rangle_3$, which can be accomplished either via geometrical protocol \cite{braid6,braid7} or magic state distillation \cite{magic}, a $\pi/8$-gate ($T_{1(2)}$ gate) can be implemented through

\begin{eqnarray}
T_1 &=& P^{s_1,s_2}_1\Pi_{3,z}^{s_3} \Pi_{3,x}^{s_2} \Pi_{13,zz}^{s_1} \;, \nonumber \\
T_2 &=& P^{s_1,s_2}_2\Pi_{3,z}^{s_3} \Pi_{3,x}^{s_2} \Pi_{23,zz}^{s_1}\;, \nonumber \\
P^{s_1,s_2}_j &=& \begin{cases}
1 & \text{if $s_1=s_2=1$} \\
Z_j \equiv \Pi_{3,z}^{s_6} \Pi_{3,x}^{-s_4} \Pi_{j3,zz}^{s_5} \Pi_{3,x}^{s_4} & \text{if $s_1=-s_2=1$} \\
\mathcal{P}_j & \text{if $s_1=-s_2=-1$} \\
\mathcal{P}_j Z_j \equiv \mathcal{P}_j \Pi_{3,z}^{s_6} \Pi_{3,x}^{-s_4} \Pi_{j3,zz}^{s_5} \Pi_{3,x}^{s_4} & \text{if $s_1=s_2=-1$}
\end{cases}
\end{eqnarray}
at the cost of turning the ancilla state to a $\sigma_z^{(3)}$ eigenstate. This can be verified by applying directly the three projectors on $|\Psi\rangle=|\psi\rangle_1|\phi\rangle_2 |M\rangle_3$. Without loss of generality, we will verify $T_1$ below by taking $|\psi\rangle_1 = \alpha |0\rangle_1 +\beta |1\rangle_1$ and showing that it transforms to $|\psi'\rangle_1 = \alpha e^{-\mathrm{i}\pi/8} |0\rangle_1 +\beta e^{\mathrm{i}\pi/8}|1\rangle_1$. To this end, we will suppress $|\phi\rangle_2$ and focus on the action of the three projectors $\Pi_{3,z}^{s_3}$, $\Pi_{3,x}^{s_2}$, $\Pi_{13,zz}^{s_1}$ on $|\psi\rangle_1|M\rangle_3$,

\begin{eqnarray}
\Pi_{3,z}^{s_3} \Pi_{3,x}^{s_2} \Pi_{13,zz}^{s_1}|\psi_1\rangle_1|M\rangle_3 &\propto& \Pi_{3,z}^{s_3} \Pi_{3,x}^{s_2} \left[\left(\alpha e^{-\mathrm{i}\pi/8} |0\rangle_1 |0\rangle_3 +\beta e^{\mathrm{i}\pi/8} |1\rangle_1 |1\rangle_3\right)\times (1+s_1)\right. \nonumber \\
&& +\left. \left(\alpha e^{\mathrm{i}\pi/8} |0\rangle_1 |1\rangle_3 +\beta e^{-\mathrm{i}\pi/8} |1\rangle_1 |0\rangle_3\right)\times (1-s_1) \right] \nonumber \\
&\propto&  \Pi_{3,z}^{s_3} \left[\left(\alpha e^{-\mathrm{i}\pi/8} |0\rangle_1 |+\rangle_3 +\beta e^{\mathrm{i}\pi/8} |1\rangle_1 |+\rangle_3\right)\times (1+s_1)\times (1+s_2)\right. \nonumber \\
&& + \left(\alpha e^{-\mathrm{i}\pi/8} |0\rangle_1 |-\rangle_3 -\beta e^{\mathrm{i}\pi/8} |1\rangle_1 |-\rangle_3\right)\times (1+s_1)\times (1-s_2)
\nonumber \\
&& + \left(\alpha e^{\mathrm{i}\pi/8} |0\rangle_1 |+\rangle_3 +\beta e^{-\mathrm{i}\pi/8} |1\rangle_1 |+\rangle_3\right)\times (1-s_1)\times (1+s_2)
\nonumber \\
&& +\left. \left(-\alpha e^{\mathrm{i}\pi/8} |0\rangle_1 |-\rangle_3 +\beta e^{-\mathrm{i}\pi/8} |1\rangle_1 |-\rangle_3\right)\times (1-s_1) \times (1-s_2) \right] \nonumber \\
&\propto& \left(\alpha e^{-\mathrm{i}\pi/8} |0\rangle_1 |0\rangle_3 +\beta e^{\mathrm{i}\pi/8} |1\rangle_1 |0\rangle_3\right)\times (1+s_1)\times (1+s_2)\times (1+s_3) \nonumber \\
&& +\left(\alpha e^{-\mathrm{i}\pi/8} |0\rangle_1 |1\rangle_3 +\beta e^{\mathrm{i}\pi/8} |1\rangle_1 |1\rangle_3\right)\times (1+s_1)\times (1+s_2)\times (1-s_3) \nonumber \\
&& +\left(\alpha e^{-\mathrm{i}\pi/8} |0\rangle_1 |0\rangle_3 -\beta e^{\mathrm{i}\pi/8} |1\rangle_1 |0\rangle_3\right)\times (1+s_1)\times (1-s_2)\times (1+s_3) \nonumber \\
&& -\left(\alpha e^{-\mathrm{i}\pi/8} |0\rangle_1 |1\rangle_3 -\beta e^{\mathrm{i}\pi/8} |1\rangle_1 |1\rangle_3\right)\times (1+s_1)\times (1-s_2)\times (1-s_3) \nonumber \\
&& +\left(\alpha e^{\mathrm{i}\pi/8} |0\rangle_1 |0\rangle_3 +\beta e^{-\mathrm{i}\pi/8} |1\rangle_1 |0\rangle_3\right)\times (1-s_1)\times (1+s_2)\times (1+s_3) \nonumber \\
&& +\left(\alpha e^{\mathrm{i}\pi/8} |0\rangle_1 |1\rangle_3 +\beta e^{-\mathrm{i}\pi/8} |1\rangle_1 |1\rangle_3\right)\times (1-s_1)\times (1+s_2)\times (1-s_3) \nonumber \\
&& +\left(-\alpha e^{\mathrm{i}\pi/8} |0\rangle_1 |0\rangle_3 +\beta e^{-\mathrm{i}\pi/8} |1\rangle_1 |0\rangle_3\right)\times (1-s_1)\times (1-s_2)\times (1+s_3) \nonumber \\
&& +\left(\alpha e^{\mathrm{i}\pi/8} |0\rangle_1 |1\rangle_3 -\beta e^{-\mathrm{i}\pi/8} |1\rangle_1 |1\rangle_3\right)\times (1-s_1)\times (1-s_2)\times (1-s_3) \;.\nonumber \\
\end{eqnarray}
It then follows that $T_1$ is automatically achieved for $s_1=s_2=1$. If $s_1=-s_2=1$, a $Z$-gate can be applied via forced measurements to flip the sign of $|1\rangle_1$. If $s_1=-s_2=-1$, a phase gate $\mathcal{P}_1$ can be applied to flip the sign of the exponentials $e^{\pm\mathrm{i}\pi/8}$. Finally, if $s_1=s_2=-1$, both $Z$-gate and $\mathcal{P}_1$ are applied.





\begin{thebibliography}{99}
\bibitem{note} By the conservation of total fermion parity, i.e., $\gamma_{0,1}\gamma_{0,2}\gamma_{0,3}\gamma_{0,4}\gamma_{\pi,1}\gamma_{\pi,2}\gamma_{\pi,3}\gamma_{\pi,4}=\pm 1$, the qubit $\sigma_z^{(3)}$ can be equivalently expressed as $\gamma_{\pi,1}\gamma_{\pi,2}\gamma_{\pi,3}\gamma_{\pi,4}$.
\end{thebibliography}
\end{document}